\documentclass{article}%
\usepackage{amsmath}
\usepackage{amsfonts}
\usepackage{amssymb}
\usepackage{graphicx}
\usepackage{epstopdf}%
\newcommand{\ba}{\begin{array}}
\newcommand{\ea}{\end{array}}

\def\Dsl[#1]#2{#1\hskip-#2pt/} 

\newcommand{\be}[1]{ \begin{eqnarray}\label{#1}}
\newcommand{\ee}{\end{eqnarray}}
\newcommand{\bea}{\begin{eqnarray}}
\newcommand{\eea}{\end{eqnarray}}

\newcommand{\ns}{\Dsl[n]{5}}
\newcommand {\nbs}{\Dsl[\bar n]{5}}
\newcommand{\nbn}{\frac{\nbs\ns}{4}}

\newcommand{\KLL}{K_{\text{LL}} }
\newcommand{\KLS}{K_{\text{LS}} }

\begin{document}

\noindent


\begin{center}
{\Large  QCD radiative corrections to the soft spectator contribution in the wide angle Compton scattering }

\vspace*{1cm}

N. Kivel\footnote{
On leave of absence from St.~Petersburg Nuclear Physics Institute,
188350, Gatchina, Russia} and  
M. Vanderhaeghen
\\[3mm]
 { \it Helmholtz Institut Mainz, Johannes Gutenberg-Universit\"at, D-55099 Mainz, Germany
 \\
 Institut f\"ur Kernphysik, Johannes Gutenberg-Universit\"at, D-55099 Mainz, Germany
 }

\vspace*{1cm}
\end{center}

\begin{abstract}
We derive  the  complete factorization formula for the leading power contribution in wide angle Compton scattering. 
It consists of the soft- and hard-spectator contributions. The hard-spectator contribution is well known 
and defined in the form of the convolution of a hard kernel with the nucleon distribution amplitudes.  The soft-spectator contribution
describes the scattering which involves the soft modes.  We use the soft collinear effective theory   in order to define this term in a field theoretical approach. 
Using the SCET  framework we provide the proof of the factorization formula.  We also compute the next-to-leading QCD corrections to the hard 
coefficient function of the soft spectator contribution and perform a  phenomenological analysis of existing experimental data within the developed formalism.

\end{abstract}

\vspace*{1cm}

\section*{Introduction}

The Wide Angle Compton scattering (WACS)  provides  an excellent possibility  to study the complicated hadronic dynamics in  hard exclusive processes.  The asymptotic behavior 
of the cross section for this process at large energy and momentum  transfer  $s\sim-t\sim -u\gg \Lambda$, where $\Lambda$ is the typical hadronic scale, 
 was predicted long time ago  using the   QCD  power counting arguments  \cite{Brodsky:1973kr, Matveev:1972gb}.  It was suggested   that 
\be{sgm-asy}
\frac{d\sigma^{\gamma p\rightarrow \gamma p}}{dt} \sim \frac 1{s^{6}} f(\theta),
\ee
where $\theta$ denotes the scattering angle in the center-of-mass  frame.  Later  the QCD factorization approach  was developed in order  to compute  the function $f(\theta)$, see {\it e.g.} Refs.\cite{Lepage:1980fj, Chernyak:1983ej} .  The asymptotic expression for this function  is dominated by the hard two-gluon exchange diagrams.  The non-perturbative dynamics in this case is described by the so-called nucleon distribution amplitude (DA)  which  is related with the  three-quark Fock component of the nucleon wave function in the light-cone formalism 
\cite{Brodsky:1989pv}. 

The first measurements of the differential cross section were carried out in a Cornell experiment \cite{Shupe:1979vg}.  It was found that the cross section displays a scaling behavior which is in a reasonable agreement with pQCD predictions (\ref{sgm-asy}). However  theoretical  estimates show   that the absolute value of the cross section  computed within the  QCD factorization framework  is much smaller than the corresponding experimental values  \cite{Kronfeld:1991kp,Vanderhaeghen:1997my, Brooks:2000nb, Thomson:2006ny}.   The new experiments carried out at JLab  for  large angle scattering  provided  more precise data for the cross section  for the different energies \cite{Danagoulian:2007gs}. The new data have much better accuracy and allow one to carry a more detailed comparison with the theoretical predictions for the cross section.   Moreover,  for the first time the  longitudinal and transverse components of the recoil proton polarization were also measured \cite{Hamilton:2004fq}.  These measurements provided  that the value of the longitudinal  asymmetry $\KLL$ is qualitatively different from the one that can be obtained in the hard-spectator  (hard two-gluon exchange)  factorization picture described above. 

All these observations  indicate that the existing data are still far  away  from the asymptotic region where the formula (\ref{sgm-asy}) based on the hard two-gluon exchange is  dominant. 
Therefore  in order to explain  the confrontation of the theory with the data one has to implement  a different  picture of the scattering which describes also  the dominant preasymptotic effects. 
 To this extent,  promising  results have been obtained 
within the hand-bag model approach  \cite{Radyushkin:1998rt,Diehl:1998kh,Diehl:1999tr,Diehl:2002ee}. 
The main idea of this model is that the dominant contribution in the relevant  kinematical region is provided by the soft-overlap contribution  which is also known as the Feynman mechanism. 
In this model both photons interact  with a single quark which couples to the soft spectators through the generalized parton distributions (GPDs).  The GPDs describe the nonperturbative soft-overlap mechanism at small momentum transfer $-t\sim \Lambda^{2}$ and their models  can be constrained  by using their relations  to  the form factors and usual parton distributions, see e.g. 
\cite{Goeke:2001tz,Diehl:2004cx,Belitsky:2005qn} and references therein. The extension of the GPD formalism  for the description of the region with large $-t\gg \Lambda$  is the key  assumption of the handbag model.  There is also an alternative approach based on the constituent quark model \cite{Miller:2004rc}.  All these models provide a satisfactory  description of  the cross section   data  and asymmetry $\KLL$ that  demonstrates  a strong support for the crucial  role played by the soft-spectator scattering  in WACS and possibly in  other  hard reactions.  

The further and more accurate  experimental studies of the WACS cross sections and asymmetries can be performed at JLab after the 12 GeV Upgrade \cite{proposal}. 
This provides a strong motivation to develop a systematic theoretical approach which accommodates consistently  both hard- and soft-spectator reaction 
mechanisms and allows one to reduce the model dependence in the data analysis.   

Some steps in this direction were made in the framework of the GPD handbag model in Refs.\cite{Diehl:2004cx,Huang:2001ej}. However in the latter framework one is faced with the  problem how to consistently perform  the matching between the  hard and the soft regions: it is not clear how to map   systematically the  infrared  poles  arising from the partonic diagrams with the ultraviolet structure of the GPD-based matrix elements.  This difficulty does not allow  to formulate a consistent theoretical  approach.

 The importance of the soft modes  and their specific role for the description of  hard processes  with  nucleons has already been observed  a long time ago in Refs. \cite{Duncan:1979hi,Duncan:1979ny}. 
 A substantial progress in the  theoretical description of such configurations has been achieved during past decade in the framework of the soft collinear effective theory (SCET) 
 \cite{Bauer:2000ew, Bauer2000, Bauer:2001ct, Bauer2001, Beneke:2002ph, Beneke:2002ni}. 
An application of this technique for the description  of the soft-overlap contribution in  different hadronic hard exclusive reactions including WACS  has recently been studied in 
Refs.\cite{Kivel:2010ns, Kivel:2012mf, Kivel:2012vs, Kivel:2013fja}.   
  The attractive feature of this  approach is the possibility to  formulate the  factorization of  the hard- and the soft-spectator contributions  consistently  using  a field theory technique. This can be done because of  the  presence of the collinear and soft modes in the effective Lagrangian.  This allows one to establish a systematic power counting rule for  all contributions which can be relevant  for a given process, and to include  the hard- and  soft-spectator  configurations on the same footing .

The soft modes in  the SCET approach  describe   particles with the momenta $k_{s}^{\mu}\sim \Lambda$. The presence of such modes unavoidably introduces the additional intermediate scale of order $\Lambda Q$ which is known as a hard-collinear scale. Such intermediate virtualities  naturally  arises due to the interactions between the collinear and soft particles. In such case the factorization of an amplitude  can be described in terms of the two following steps.  In a first step one integrates out the hard modes describing the particles with momenta $p_{h}^{\mu}\sim Q$.  The remaining relevant degrees of  freedom are then described by the hard-collinear, collinear and soft particles.  If the value of the hard scale $Q$  is quite large so that  the hard-collinear scale $\mu_{hc}\sim\sqrt{\Lambda Q}$  is  also quite large,  then one can proceed further and to integrate over the hard-collinear modes. After that  the unknown nonperturbative dynamics  is described by  matrix elements of  the operators constructed only from  collinear or soft fields.  In case  the hard-collinear scale is not large, for instance $\mu_{hc}\lesssim 1$GeV then only the first step factorization can be performed.  
Notice that such a framework  provides a clear theoretical understanding  of the intermediate region of $Q^{2}$.  Namely,  this region can be associated with the values of $Q^{2}$  for which the hard-collinear scale is not large enough and the full two-step factorization description  can not provide an accurate description.  It seems that  this situation is relevant in the  description of many hard exclusive processes  including also WACS which is the subject of  this paper.   

The  complete  QCD factorization formula which includes the hard- and soft-spectator contributions  for  WACS has been suggested in Ref.\cite{Kivel:2012vs}. It was shown there that the soft-spectator configuration in WACS and elastic electron-proton scattering  is described by the same matrix element and this allows to constrain a particular  contribution entering  the two-photon exchange amplitudes.  However in Ref.\cite{Kivel:2012vs}  the factorization for  Compton scattering  was not discussed in detail.  In the current publication we provide the details of this formalism.  We provide the proof of the factorization formula within the SCET framework and  also compute the next-to-leading hard corrections to coefficient functions in front of the SCET operator which describe the soft-spectator contribution.  Then  we use  the obtained results  in the phenomenological analysis of existing data.  

Our paper is organized as follows.  In Section~\ref{wide} we  discuss the general properties  of the WACS.   We specify notations and kinematics, discuss the amplitude and  cross section. 
In the Section~\ref{factorization} we discuss the QCD factorization of the WACS amplitudes within  the SCET framework. We prove that the complete factorization formula consists of two contributions associated with the soft- and hard-spectator scattering and provide their theoretical description.  Section~\ref{calculation} is devoted to the calculation of the next-to-leading  order coefficient functions for the soft-spectator contribution. 
In  Section~\ref{phenomenology} we use the obtained expressions for the analysis of existing experimental data. In Appendix we present more  details about the next-to-leading calculations.

\section{  Wide angle  Compton scattering: general remarks }
\label{wide}

The kinematics of the real Compton scattering $\gamma(q)+N(p)\rightarrow \gamma(q')+N(p')$ is described by the Mandelstam variables
\be{Mvar}
s=(p+q)^{2},\quad t=(p'-p)^{2}<0,\quad u=(p-q')^{2}<0, \quad s+t+u=2m^{2}.
\ee
In the center-of-mass frame $\vec{p}+\vec{q}=0$ the particle momenta read
\bea
p= E_{cm}\left(\sqrt{1+\frac{m^{2}}{E_{cm}^{2}} },0,0,-1\right),~\ q=E_{cm}(1,0,0,1),
\\
p^{\prime}= E_{cm}\left(\sqrt{1+\frac{m^{2}}{E^{2}} },-\sin\theta,0,-\cos\theta\right),
~\ q^{\prime}=E_{cm}(1,\sin\theta,0,\cos\theta),
\eea
where $m$ denotes the nucleon mass and  $\theta$ is the scattering angle in the center-of-mass frame.  

In what follows,  we will also use 
the  auxiliary light-cone vectors $n,\  \bar n$ such that 
\be{def:lc}
 n=(1,0,0,-1),~\bar{n}=(1,0,0,1),~\ (n\cdot \bar{n})=2.
\ee
  In order to write the light-cone expansions of the particle momenta we  choose a frame where $z$-axis is directed along the incoming and outgoing  nucleons. 
 For  large-angle kinematics $s\sim-t\sim-u\gg m^{2}\sim \Lambda^{2}$  one obtains 
 \bea
p&=&\sqrt{-t}\frac{\bar{n}}{2}+\frac{m^{2}}{\sqrt{-t}}\frac{n}{2}\simeq \sqrt{-t}\frac{\bar{n}}{2},\quad
p^{\prime}=\frac{m^{2}}{\sqrt{-t}}\frac{\bar{n}}{2}+\sqrt{-t}\frac{n}{2}\simeq \sqrt{-t}\frac{n}{2}, 
\label{mom1},
\\
q&\simeq&  \frac{u}{t}\  p-\frac{s}{t}\ p'+q_{\bot},~
\ q^{\prime}\simeq -\frac{s}{t}\ p+ \frac{u}{t}\  p'+q_{\bot},
\eea
where we neglected the power corrections $m/Q$  and assume $-t\simeq\frac12 s(1-\cos\theta)$, $-u\simeq \frac12 s(1+\cos\theta)$.  
 
In order to describe the  amplitude  of the Compton scattering  it is convenient to introduce the following notations
 \cite{Babusci:1998ww}
\bea
P=\frac{1}{2}(p+p^{\prime}),~K=\frac{1}{2}(q+q^{\prime}),~P^{\prime}=P-K\frac{(P\cdot K)}{K^{2}},
\\
N_{\mu}=\varepsilon_{\mu\alpha\beta\gamma}P^{\alpha}\frac{1}%
{2}(p-p^{\prime})^{\beta}K^{\gamma},~\ \varepsilon_{0123}=+1,
\eea
Then the real  Compton amplitude on the proton  can be described  in terms of six scalar amplitudes $T_{i}(s,t)$
\be{Smat}
\langle p',q'; out| S-1|in; p,q\rangle=i(2\pi)^{4}\delta(p'+q'-p-q) M^{\gamma p\rightarrow \gamma p}.
\ee
with
\bea
M^{\gamma p\rightarrow \gamma p}&=&-e^{2}\,  \varepsilon^{\ast\mu}( q^{\prime})\varepsilon^{\nu}(q) 
\bar{N}(p^{\prime})\left\{-\mathcal{T}_{12}^{\mu\nu}\left(  T_{1}+\Dsl[K]{7}~T_{2}\right)  -\mathcal{T}_{34}^{\mu\nu
}\left( T_{3}+\Dsl[K]{7}~T_{4}\right)
\right. \nonumber \\  
&& \left.
\phantom{ \varepsilon^{\ast\mu}( q^{\prime})\varepsilon^{\nu}(q)  \bar{N}(p^{\prime}) \{-\mathcal{T}_{12} } 
 +\mathcal{T}_{5}^{\mu\nu}i\gamma_{5}~T_{5}%
+\mathcal{T}_{6}^{\mu\nu}~i\gamma_{5}\Dsl[K]{7}~T_{6}\right\} N(p),
\label{defM}
\eea
where  $e$ denotes the electromagnetic charge of the proton.  In Eq.(\ref{defM}) we introduced  the following tensor  structures
\begin{align}
  \mathcal{T}_{12}^{\mu\nu}=-\frac{P^{\prime\mu}P^{\prime\nu}}{P^{\prime2}},\quad   
 \mathcal{T}_{34}^{\mu\nu}=\frac{N^{\mu}N^{\nu}}{N^{2}},\quad
  \mathcal{T}_{5}^{\mu\nu}=\frac{P^{\prime\mu}N^{\nu}-P^{\prime\nu}N^{\mu}}{P^{\prime2}K^{2}},\quad
\mathcal{T}_{6}^{\mu\nu}=\frac{P^{\prime\mu}N^{\nu}+P^{\prime\nu}N^{\mu}}{P^{\prime2}K^{2}}.
 \label{T1-6}
\end{align}%
One can easily see that the tensor $\mathcal{T}_{j\ \mu\nu}$ are gauge invariant and orthogonal
\bea
&   q^{\nu}\mathcal{T}_{j\ \mu\nu}= q'^{\mu}\mathcal{T}_{j\ \mu\nu}=0,\quad    \mathcal{T}_{i}^{\mu\nu}\mathcal{T}_{j\ \mu\nu}=0,\quad i\neq j, 
\label{TiTj} &\\
& \mathcal{T}_{12}^{\mu\nu}\mathcal{T}_{12\mu\nu}= \mathcal{T}_{34}^{\mu\nu}\mathcal{T}_{34\mu\nu}=1, \quad
 \mathcal{T}_{5}^{\mu\nu}\mathcal{T}_{5\mu\nu}= \mathcal{T}_{6}^{\mu\nu}\mathcal{T}_{6\mu\nu}=2.&
 \label{TiTi}
\eea
This allows one  to compute  the contribution to a given amplitude using appropriate contractions.\footnote{ 
The parametrization (\ref{defM}) was introduced for description of the low-energy scattering in Ref.\cite{Babusci:1998ww}. 
Therefore one can find that such definitions are not convenient for the analysis
of the large energy behavior because the scalar amplitudes $T_{i}$ are not dimensionless. 
This can be cured by the appropriate redefinition of the amplitudes but we decided to keep the original notations. }

 The unpolarized differential  cross section describing  real Compton scattering  reads \cite{Babusci:1998ww}
\bea
\frac{d\sigma}{dt}&=&\frac{\pi\alpha^{2}}{(s-m^{2})^{2}} 
\left\{
(s-m^{2})(m^{2}-u)\frac{1}{2} (|T_{2}|^{2}+|T_{4}|^{2})+(m^{4}-su)|T_{6}|^{2}
\right. \nonumber\\
&& \left. +m(s-u)\operatorname{Re}\left[  T_{1}T_{2}^{\ast}+T_{3}T_{4}^{\ast}\right]
+\frac{1}{2}(4m^{2}-t)(|T_{1}|^{2}+|T_{3}|^{2})-t|T_{5}|^{2}
\right\} , 
\label{dsig}
\eea
where the fine structure coupling $\alpha=e^{2}/4\pi\simeq 1/137$. The explicit expressions for the  observables including the longitudinal polarization  $\KLL $ can  also be found in   Ref.\cite{Babusci:1998ww}  and we will not rewrite it here.

\section{ QCD factorization for wide angle Compton scattering within the SCET framework}
\label{factorization}

\subsection{ SCET:  preliminary remarks}

In this section we consider in detail the leading power  factorization of the WACS amplitudes 
 using  the SCET approach developed in Refs.\cite{Bauer:2000ew, Bauer2000, Bauer:2001ct, Bauer2001, Beneke:2002ph, Beneke:2002ni}. 
 Below we assume  that  the  relevant dominant  regions are described by  particles which 
 have hard  $p_{h}$, hard-collinear $p_{hc}$, collinear $p_{c}$ and soft $p_{s}$ momenta.
The light-cone components $\left(  p\cdot n,p\cdot \bar{n},p_{\bot}\right)  \equiv
(p_{+},p_{-},p_{\bot})$ of the corresponding momenta scale as
\begin{equation}
p_{h}\sim Q\left(  1,1,1\right)  ,~p_{h}^{2}\sim Q^{2}, \label{def:ph}%
\end{equation}%
\begin{equation}
p_{hc}\sim Q\left(  1,\lambda^{2},\lambda\right)  \text{ or }p_{hc}^{\prime
}\sim Q\left(  \lambda^{2},1,\lambda\right)  ,~p_{hc}^{2}\sim Q^{2}\lambda
^{2}\ \sim Q\Lambda, \label{def:phc}%
\end{equation}%
\begin{equation}
p_{c}\sim Q\left(  1,\lambda^{4},\lambda^{2}\right)  \text{ or ~}p_{c}%
^{\prime}\sim Q\left(  \lambda^{4},1,\lambda^{2}\right)  ,~\ p_{c}^{2}\sim
Q^{2}\lambda^{4}\sim\Lambda^{2}, \label{def:pc}%
\end{equation}%
\begin{equation}
p_{s}\sim Q\left(  \lambda^{2},\lambda^{2},\lambda^{2}\right)  ,~\ p_{s}%
^{2}\sim Q^{2}\lambda^{4}\sim\Lambda^{2}. \label{ps}%
\end{equation}
The scales    $Q$ and $\Lambda$ denote the generic large and soft scales, respectively.
The small dimensionless parameter $\lambda$ is set to be  $\lambda\sim\sqrt{\Lambda/Q}$. 
Let us assume  that there are no other relevant  modes required for the factorization of the
leading power amplitudes. In this case the factorization can be described in
two steps: first, we integrate out the hard modes and reduce the full QCD to the
effective theory. The corresponding effective Lagrangian is constructed
from the hard-collinear and soft particles. This effective theory
is denoted as SCET-I. If the hard scale $Q$ is so large that the hard
collinear scale $\mu_{hc}\sim\sqrt{Q\Lambda}$ is a good parameter for the
perturbative expansion then one can perform the second matching step and
to integrate out the hard-collinear modes. Then the resulting effective
Lagrangian is constructed only from the collinear and soft fields and
the corresponding effective theory is denoted as SCET-II.

For the SCET fields we use the following notations. The fields $\xi_{n}^{C},$
$A_{\mu C}^{(n)}$ and $\xi_{\bar{n}}^{C},~A_{\mu C}^{(\bar{n})}~$ denote the
hard-collinear ($C=hc$) or collinear ($C=c$) quark and gluon fields associated
with momentum $p^{\prime}$ and $p$, respectively, see Eq.(\ref{mom1}). As
usually, the hard-collinear and collinear quark fields satisfy 
\begin{equation}
\Dsl[n]{5.5} \xi_{n}^{C}=0,~~
\Dsl[\bar n]{5.5}  \xi_{\bar{n}}^{C}=0.
\end{equation}
The fields $q$ and $A_{\mu}^{(s)}$ denote the soft quarks and gluons with the
soft momenta as in Eq.(\ref{ps}) which also enter in the SCET Lagrangian.

In the wide-angle kinematics we have the energetic particles propagating with
large energies in four directions. Therefore it is useful to introduce two
more auxiliary light-cone vectors associated with the photon momenta $q$ and
$q^{\prime}$%
\begin{equation}
\bar{v}^{\mu}=\frac{2q^{\mu}}{\sqrt{-t}},~\ v^{\mu}=\frac{2q^{\prime\mu}%
}{\sqrt{-t}},~(\bar{v}\cdot v)=2.
\end{equation}
Using the vectors $\bar{v},v$ we also introduce the hard-collinear quark and
gluon fields in the similar way as before just changing $(n,\bar
{n})\rightarrow\left(  v,\bar{v}\right)  $.

The explicit expression for the SCET-I~ Largangian in position space was derived
 in Refs.\cite{Beneke:2002ph, Beneke:2002ni}. This Lagrangian can be represented  in the form of expansion with respect to the
small parameter $\lambda$
\begin{equation}
\mathcal{L}_{\text{SCET}}^{(n)}=\mathcal{L}_{\xi\xi}^{(0,n)}+\mathcal{L}_{\xi\xi}^{(1,n)}+\mathcal{L}_{\xi q}^{(1,n)}+\mathcal{O}(\lambda^{2}),
\label{Lscet}%
\end{equation}
where $\mathcal{L}^{(\lambda,n)}\sim\mathcal{O}(\lambda)$. The 
explicit expressions for the simplest terms read
\begin{equation}
\mathcal{L}_{\xi\xi}^{(0,n)}=\bar{\xi}_{n}^{hc}(x)\left(  i~n\cdot D+gn\cdot
A^{(s)}(x_{-})+i\Dsl[D]{7}_{\bot}\frac{1}{i\bar{n}\cdot D}i\Dsl[D]{7}_{\bot}\right)  \frac
{\bar{n}}{2}\xi_{n}^{hc}(x), \label{Ln0}%
\end{equation}%
\begin{equation}
\mathcal{L}_{\xi q}^{(1,n)}=\bar{\xi}_{n}^{hc}(x)i\Dsl[D]{7}_{\bot}W_{n}~q(x_{-})+\bar
{q}(x_{-})W_{n}^{\dag}i\Dsl[D]{7}_{\bot}\xi_{n}^{hc}(x). \label{Ln1}%
\end{equation}
The expression for  the subleading term $\mathcal{L}_{\xi\xi
}^{(1,n)}$ is a bit lengthy and we will not rewrite it here.  The similar
expressions are also valid for the other collinear sectors associated with the
directions $\bar{n},v,\bar{v}$.   In  the above formulas  we used  
$iD_{\mu}=i\partial_{\mu}+gA_{\mu hc }^{(n)}$,~$x_{-}=\frac{1}{2}(x\cdot\bar{n})n$ and the hard-collinear Wilson lines
\begin{equation}
W_{n}(z)=\text{P}\exp\left\{  ig\int_{-\infty}^{0}ds~\bar{n}\cdot A_{hc}^{(n)}(z+s\bar{n})\right\}  .
\end{equation}

The matching from SCET-I to SCET-II  can be performed by integration over the hard-collinear fields. Technically this can be done via the substitutions 
 $\xi_{n}^{hc}\rightarrow\xi_{n}^{c}+\xi_{n}^{hc}$ and 
$A_{hc}^{(n)}\rightarrow  A_{c}^{(n)}+A_{hc}^{(n)}$ in the Lagrangian (\ref{Lscet}) and the external SCET-I operators  and integration over the  hard-collinear
fields.  Following  this way one has to deal with  the intermediate theory which includes the hard-collinear, collinear and soft fields. 
A more detailed description of this step can be found in  Refs.\cite{Bauer:2002nz,Bauer:2003mga} in the hybrid formulation  and in
Ref.\cite{Beneke:2003pa} in the position space formulation. 

The  power counting rules can be fixed using the  power counting of the SCET fields. These  rules for the SCET fields can
be obtained from the corresponding propagators in momentum space and read (see,  for instance,  Ref.\cite{Beneke:2002ph})
\begin{equation}
\xi_{n}^{hc}\sim\lambda,~\ \bar{n}\cdot A_{hc}^{(n)}\sim1,~A_{\bot hc}%
^{(n)}\sim\lambda~\ ,n\cdot A_{hc}^{(n)}\sim\lambda^{2},
\end{equation}%
\begin{equation}
\xi_{n}^{c}\sim\lambda^{2},~\ \bar{n}\cdot A_{c}^{(n)}\sim1,~A_{\bot c}%
^{(n)}\sim\lambda^{2}~\ ,n\cdot A_{c}^{(n)}\sim\lambda^{4}, \label{colf}%
\end{equation}%
\begin{equation}
A_{s}^{\mu}\sim\lambda^{2},~\ q\sim\lambda^{3}.
\end{equation}

Performing the matching from QCD to SCET-I one has to consider  different operators built from the SCET-I fields.   
It is convenient  to construct  such operators  using  the following  gauge invariant combinations
\begin{equation}
\chi_{n}^{hc}(\lambda\bar{n})\equiv~W_{n}^{\dag}(\lambda\bar{n})\xi_{n}%
^{hc}(\lambda\bar{n}),~\ \bar{\chi}_{n}^{hc}(\lambda\bar{n})\equiv\bar{\xi}%
_{n}^{hc}(\lambda\bar{n})W_{n}(\lambda\bar{n}), \label{Chi}%
\end{equation}%
\begin{equation}
\mathcal{A}_{\mu\ hc}^{(n)}(\lambda\bar{n})\equiv\left[  W_{n}^{\dag}%
(\lambda\bar{n})D_{\mu }W_{n}(\lambda\bar{n})\right]  , \label{AC}%
\end{equation}
where the covariant derivative $D_{\mu}$
is applied only  inside the brackets.  For  collinear operators one can use  similar expressions. 

\subsection{ Factorization for the WACS amplitudes }

The   factorization formula for the WACS amplitudes can be
written as a sum of  two contributions describing the soft- and the hard-spectator 
scattering. The hard-spectator scattering contribution  dominates at  asymptotically large 
values of $Q\rightarrow \infty$.  But  for the moderate values of $Q$, where the hard-collinear virtualities $Q\Lambda\lesssim m^{2}$ are not large,
one has to take into account the soft-overlap contribution which, in general,  
is described by the matrix elements of the all appropriate  SCET-I  operators.   

\begin{figure}[ptb]%
\centering
\includegraphics[width=4.0in]{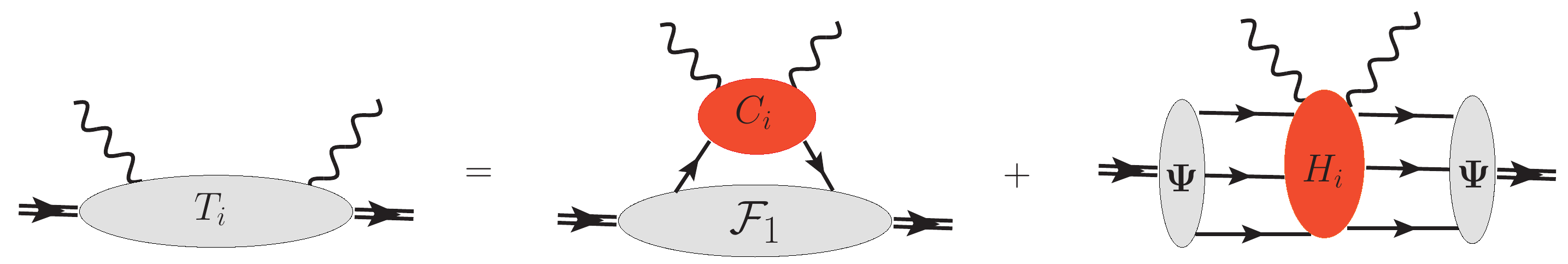}%
\caption{The reduced diagrams describing the factorization formula in Eq.(\ref{Ai-fact}). }%
\label{wacs-scet}%
\end{figure}

Below we show  that the  complete leading power factorization formula can be written as 
\begin{equation}
T_{i}(s,t)=C_{i}(s,t)\mathcal{F}_{1}(t)+ \mathbf{\Psi}\ast H_{i}(s,t)\ast\ \mathbf{\Psi},\quad  i=2,3,4. 
\label{Ai-fact}  
\end{equation}
Here the first  term describes the soft-spectator contribution while the second term corresponds to the well known hard-spectator  mechanism. 
 For illustration, in Fig.\ref{wacs-scet}  we show the different contributions in Eq.(\ref{Ai-fact})  as appropriate reduced diagrams.  
 In the large $Q$ limit  {\it both} contributions in   Eq.(\ref{Ai-fact})  behave as $Q^{-4}$ up to logarithmic corrections and give 
 \be{Tias}
 T_{2,4,6}(s,t)\sim Q^{-4},\quad    Q\rightarrow \infty.
 \ee
For large values of $Q$ the soft-spectator  scattering is strongly suppressed due to the so-called Sudakov logarithms and therefore the
 hard-spectator contribution  becomes dominant. But for moderate values of $Q$ the  effect of the Sudakov suppression is still weak, see {\it e.g.} discussion in Ref.\cite{Kivel:2012mf}, 
 therefore  the soft-spectator contribution is quite large or even dominant.    

The expression in Eq.(\ref{Ai-fact})  does not include  the amplitudes $T_{1,3,5}$.  These amplitudes   describe the  scattering  when the helicity of nucleon 
 is not conserved.  These amplitudes are suppressed as $\mathcal{O}(1/Q^{5})$ and their 
factorization is described by the subleading operators in SCET.   Therefore we  postpone the study of these contributions to future publications.    

The asterisks in Eq.(\ref{Ai-fact}) denote the convolution with respect to the collinear quark fractions. The non-perturbative dynamics in the hard-spectator 
contributions is described by nucleon distribution amplitudes $\mathbf{\Psi}$ which are defined by  the following matrix elements
\begin{equation}
~4\left\langle 0\left\vert
 [\chi_{\bar{n}}^{c}]^{i}_{\alpha}(\lambda_{1}n)~[\chi_{\bar{n}}^{c}]^{j}_{\beta}(\lambda_{2}n)~[\chi_{\bar{n}}^{c}]^{k}_{\sigma}(\lambda_{3}n)
 \right\vert
p\right\rangle =\frac{\varepsilon^{ijk}}{3!}\int Dx_{i}~e^{-ip_{+}\left(  \sum
x_{i}\lambda_{i}\right)  }\mathbf{\Psi}(x_{i}), 
\label{DA:def}
\end{equation}
where $p_{+}\equiv (p\cdot n) $, 
\begin{equation}
[\chi_{\bar{n}}^{c}]^{i}_{\alpha}(z)\equiv\text{P}\exp\left\{  ig\int_{-\infty}
^{0}dt~(n\cdot A_{c})(z+tn)\right\}  \left[  \xi_{\bar n}^{c}\right]_{\alpha}^{i}(z).
\end{equation}
Here the indices $i$ and $\alpha$ describe the color and Dirac indices,
respectively. The measure in Eq.(\ref{DA:def}) reads $Dx_{i}=dx_{1}dx_{2}dx_{3}\delta
(1-x_{1}-x_{2}-x_{3}).$  The function $\mathbf{\Psi}(x_{i})$ can be rewritten in terms of the three scalar amplitudes  $V, A$ and $T$ \cite{Chernyak:1983ej}
\begin{align}
 \mathbf{\Psi}(x_{i})  &  =
 V(x_{i})p_{+}\left[\frac12  \nbs C\right]_{\alpha\beta}\left[  \gamma_{5}N_{\bar{n}}\right]  _{\sigma}
+A(x_{i})p_{+}\left[  {\frac{1}{2}}\nbs\gamma_{5}C\right]  _{\alpha\beta}\left[N_{\bar{n}}\right]  _{\sigma}
+T(x_{i})p_{+}\left[\frac12  \nbs \gamma_{\bot}~C\right]  _{\alpha\beta}\left[\gamma^{\bot}\gamma_{5}N_{\bar{n}}\right]  _{\sigma}, 
\label{Psi:def}%
\end{align}
Here the large component $N_{\bar{n}}$ of the nucleon spinor is defined as
\begin{equation}
N_{\bar{n}}=\nbn N(p),
\end{equation}
and $C$ is charge conjugation matrix.  For simplicity we do not show in Eq.(\ref{DA:def}) the flavor indices.  
 
 The hard coefficient functions  $H_{i}(s,t)$  in Eq.(\ref{Ai-fact})  define the full  dependence of the hard-spectator contribution on the Mandelstam variables. 
 The leading-order approximation for these functions are defined by the two-gluon exchange  diagrams and therefore they are of order $\alpha^{2}_{s}(Q^{2})$.   

The non-perturbative dynamics in the soft-spectator contribution in Eq.(\ref{Ai-fact})  is described by the SCET form factor (FF) $\mathcal{F}_{1}(t)$.   
In the SCET framework  it  is defined as
\begin{equation}
\left\langle p^{\prime}\right\vert O^{\sigma}\left\vert p\right\rangle _{\text{{\tiny SCET}}}=\bar{N}_{n}\gamma_{\bot}^{\sigma}N_{\bar{n}%
}~\mathcal{F}_{1}(t),
\label{defF1}
\end{equation}
with the operator%
\begin{equation}
O^{\sigma}=\sum e_{q}^{2}\left\{ \bar \chi_{n}^{hc}\gamma_{\bot}^{\sigma}\chi_{\bar{n}}^{hc}
-\bar \chi_{\bar{n}}^{hc}\gamma_{\bot}^{\sigma}\chi_{n}^{hc}\right\},
\label{defO}%
\end{equation}
where the hard-collinear quark fields in the brackets $\{...\}$ have an appropriate flavor.  This SCET FF depends only from  the large momentum transfer $t$. 
However  this dependence is associated only  with the hard-collinear modes which can not be factorized for small values of the hard-collinear scale.   
One can see from  Eq.(\ref{Ai-fact}) that the energy ($s$) dependence is completely defined by the hard coefficient function $C_{i}$ and therefore
can be computed in the perturbation theory. 

The factorization  formula in  Eq.(\ref{Ai-fact}) does not  include  the contribution of the gluon SCET operator which 
is of the same order $\mathcal{O}(\lambda^{2})$  as the quark operator  $O^{\sigma}$. 
 In the next section we provide a more detailed explanation of this fact.               
             
The FF $\mathcal{F}_{1}(t)$ and the hard coefficient functions $C_{i}$ also depend on the factorization scale $\mu_{F}$ which is  not shown for simplicity. 
This scale defines the separation between the  hard and hard-collinear regions.  This scale dependence can be computed from the renormalization of the 
operator $O^{\sigma}$  (\ref{defO}) in SCET.  This provides a possibility for  systematic  calculations of the higher order corrections associated with the 
hard region.

  Performing the further factorization at very large $Q$ one can show that the SCET FF $\mathcal{F}_{1}(t)$ decreases with the same power 
as the hard spectator contribution \cite{Kivel:2010ns}
 \be{F1as}
 \mathcal{F}_{1}(t)\sim (-t)^{-2}.  
 \ee
However there is one more  subtlety which is hidden in the formal definitions and  related to the  separation of the hard and soft contributions in Eq.(\ref{Ai-fact}).  
Usually it is  accepted that the DA defined in Eq.(\ref{DA:def}) 
has an appropriate  end-point behavior which provides the convergence of the  collinear convolution integrals in  the hard-spectator contribution in Eq.(\ref{Ai-fact}). 
But a careful analysis of the  collinear and soft regions in the Feynman diagrams  allows one  to conclude that  the end-point behavior is more complicated  
and the corresponding collinear  integrals must be IR-singular in the end-point region. This IR-singularity  must cancel  with  the  UV-singularity in the soft contribution 
so that only  the whole  sum in Eq.(\ref{Ai-fact})   is well defined.  The UV-divergency  of the  FF $\mathcal{F}_{1}(t)$ can be clearly observed after 
factorization of the hard-collinear modes and transition to the SCET-II \cite{Kivel:2012mf}.    
 This  singularity is a consequence of  the overlap between  the soft and collinear regions.  Therefore in order to define  the hard and soft 
 contribution in Eq.(\ref{Ai-fact}) unambiguously one has to imply an additional,  rapidity regularization which leads to an additional  scale dependence.  
 This point is crucial  for the correct definition and calculation of the  hard-spectator term in Eq.(\ref{Ai-fact}).   Below we propose how to avoid this problem  
 using the physical subtraction scheme suggested to resolve  the similar situation in the description of $B$-decays in  Refs.\cite{Beneke:2000ry, Beneke:2000wa}. 
 
 As a last remark let us  have a closer look at the  structure of the operator $O^{\sigma}$  defined in Eq.(\ref{defO}).  It is described by  two terms which can be interpreted as 
 contribution of the quark $q$ and antiquark $\bar q$
 \bea
\mathcal{F}^{q}_{1}&=&\left\langle p^{\prime}\right\vert \bar \chi_{n}^{hc}\gamma_{\bot}^{\sigma}\chi_{\bar{n}}^{hc} \left\vert p\right\rangle _{\text{{\tiny SCET}}}, 
\\
\mathcal{F}^{\bar q}_{1} &=& \left\langle p^{\prime}\right\vert  \bar \chi_{\bar{n}}^{hc}\gamma_{\bot}^{\sigma}\chi_{n}^{hc}   \left\vert p\right\rangle _{\text{{\tiny SCET}}} .
 \eea 
It turns out that in the large $Q$ limit the contribution of the  antiquark FF is more suppressed than the  quark  one because the leading twist DA  consists  only of quark fields.
This can be seen explicitly in the matching from SCET-I to SCET-II. However the corresponding  operators  have the same hard coefficient functions  and in the region where the hard-collinear 
scale is still relatively small (the moderate  values of $Q^{2}$)  we do not have 
 strong arguments which lead to the conclusion that  the antiquark contribution is small. Therefore we included this term on the same footing as the quark contribution.  However let us notice that  the soft-collinear overlap discussed  above is associated only with the quark  form factors in  $\mathcal{F}^{q}_{1}$.

\subsection{Proof of the structure of the  leading power contribution in SCET}

The factorization of the hard modes can be described as a matching of the
T-product of the electromagnetic currents onto SCET operators. To the leading
power accuracy the operator  describing   the Eq.(\ref{Ai-fact})  can be written as
\begin{equation}
T\left\{  J_{\text{em}}^{\mu},J_{\text{em}}^{\nu}\right\}  =C^{\mu\nu\sigma
}O^{\sigma}+O_{n}^{(6)}\ast H^{\mu\nu}\ast O_{\bar{n}}^{(6)}+\mathcal{O}%
(\lambda^{13}).
\label{TJJscet}
\end{equation}
The SCET-I operator $O^{\sigma}$ is defined in Eq.(\ref{defO}).
 The  collinear operators $O_{n,\bar{n}}^{(6)}$ in Eq. (\ref{TJJscet}) are built from the three
collinear quark fields (\ref{DA:def}) or schematically
\begin{equation}
O_{n}^{(6)}=\bar{\chi}_{n}^{c}\bar{\chi}_{n}^{c}\bar{\chi}_{n}^{c}%
,~\ O_{\bar{n}}^{(6)}=\chi_{\bar{n}}^{c}\chi_{\bar{n}}^{c}\chi_{\bar{n}}^{c},
\end{equation}
where we do not write explicitly the color and spinor indices and do not show
the arguments of the fields for reasons of simplicity. The matrix element of the
each collinear operator is described by the nucleon distribution amplitudes as
in Eq.(\ref{DA:def}). The asterisks in Eq.(\ref{TJJscet}) denote the collinear
convolutions. It is easy to see from Eqs.(\ref{def:pc}) that the collinear operator $O_{n}%
^{(6)}O_{\bar{n}}^{(6)}$ is of order $\lambda^{12}$.    

The first term on the {\it rhs} of Eq.(\ref{TJJscet})   describes  the same soft-spectator scattering.   
We assume that  the SCET-I operator $O^{\sigma}\sim\mathcal{O}(\lambda^{2})$ is built from the  hard-collinear fields.  
In order to establish the behavior of this contribution at large $Q$ one has to perform the matching of this operator onto  SCET-II operators with appropriate structure.
Such  SCET-II  operator  is constructed  from the collinear and soft fields.  
The  operators built from the collinear fields  describe  the overlap with the  hadronic states as in the hard-spectator case. 
The soft operator  describes the interactions of the soft fields and  can be associated with the soft spectators.   
 One can not exclude that such  complicated configuration can  have  the same  scaling behavior of
order $ \lambda^{12}$ as 
the  hard-spectator contribution associated with the collinear operator $O_{n}^{(6)}O_{\bar{n}}^{(6)}$.  
Moreover  due to the overlap of the soft and collinear sectors  the hard- and the soft-spectator contributions can overlap,  leading to the 
end-point divergencies in the both terms.  
We suggest  that in this case  the soft-spectator configuration must be included into the factorization scheme on  the same footing as the hard-spectator term. 

In what follows we  study the contributions of the different SCET-I operators  describing the soft-overlap configuration and show that  only the operator 
defined in Eq.(\ref{defO}) provides the SCET-II operator  of order $ \lambda^{12}$.  In order to establish  this we need to consider   all possible  
 SCET-I  operators and  estimate the SCET-II operators which can be obtained from them.   

Suppose that we want  study the contribution of  the SCET-I  operator $O^{(k)}$. 
The  integration over the hard-collinear
particles  is equivalent to the calculation of the different $T$-products of the
operator $O^{(k)}$ in the intermediate SCET-I theory after substitution
\begin{equation}
\xi^{hc}\rightarrow\xi^{hc}+\xi^{c}, \  A_{\mu}^{hc}\rightarrow A_{\mu}^{hc}+A_{\mu}^{c},
\label{sub},
\end{equation}
in  all relevant collinear sectors  \cite{Bauer:2002nz, Bauer:2003mga, Beneke:2003pa}.
 The interaction vertices constructed from the hard-collinear, collinear and soft fields  are generated by the operator $O^{(k)}$ and SCET-I
Lagrangian taking into account  the substitution (\ref{sub}). 
 Computing  $T$-products one contracts the hard-collinear fields  and  obtains the SCET-II operator
\begin{equation}
T\left\{  O^{(k)},\mathcal{L}_{\text{int}}^{(l_{1},n)},\mathcal{L}%
_{\text{int}}^{(m_{1},\bar{n})},...\right\}  =O_{n}^{(\lambda_{2})}\ast J_{n}\ast
O_{S}\ast J_{\bar{n}}\ast O_{\bar{n}}^{(\lambda_{1})}\sim\mathcal{O}(\lambda
^{k+l_{1}+m_{1}+\dots}). \label{Ok-gen}%
\end{equation}
Here $\mathcal{L}_{\text{int}}^{(l_{1},n)}$ and $\mathcal{L}_{\text{int}%
}^{(m_{1},\bar{n})}$ denote the interaction vertices of order $\lambda^{l_{1}}$
and $\lambda^{m_{1}}$ associated with the $n$- and $\bar{n}$-collinear sectors,
respectively. The functions $J_{n}$ and $J_{\bar{n}}$ denote the so-called
jet-functions which describe the contractions of the hard-collinear fields,  the
asterisks in Eq.(\ref{Ok-gen}) denote the integral convolutions. 
The collinear operators $O_{\bar{n}}^{(\lambda_{1})}$ and $O_{n}^{(\lambda_{2})}$  in Eq.(\ref{Ok-gen})   describe the overlap
with the initial and final hadronic states respectively.  As before, the indices $\lambda_{1,2}$  denote  the order  of each operator.  
The soft operator $O_{S}$
is built from  the soft quark and/or  gluon fields.  
Let us refer to the operator on the  {\it rhs} of Eq.(\ref{Ok-gen})  as soft-collinear operator. 

The presence of
the soft fields in the  soft-collinear operator  allows one to associate this operator  with the soft-overlap contribution. 
In particular,   the configuration  in Eq.(\ref{Ok-gen})   can be interpreted  as  the soft-overlap of the  initial and final hadronic states. 
One can also introduce  SCET-I operators which can be associated with the soft-overlap with  the photon states.  Such operators have a more 
complicated structure and will also be considered below.

We are interested to find  all  contributions  (\ref{Ok-gen})  which provide the soft-collinear operators of  order $\lambda^{12}$ or smaller.  
It is natural to expect  that such  contributions include  the leading-order collinear operators with  $\lambda_{1}=\lambda_{2}=6$. 
The soft fields on \textit{rhs}  Eq.(\ref{Ok-gen}) increase the power of $\lambda$ but behavior of the
jet-functions can compensate this effect and therefore we can obtain the
contribution which has the same behavior as the hard-spectator one. 

Our task  is to demonstrate that the SCET-I operator $O^{\sigma}$ in
Eq.(\ref{TJJscet})  is the only  possible operator which describes
the soft-overlap contribution at the leading power accuracy. For that purpose
we  are going to study  the $T$-products  of  all possible operators $O^{(k)}$ constructed from the hard-collinear and collinear fields and  with
$k<12$.

This analysis is simpler if one takes into account that the contractions of the hard-collinear fields in  each collinear sector  can be performed independently. 
Assuming that  the SCET-I operator $O^{(k)}$ in Eq.(\ref{Ok-gen})  is built from the hard-collinear fields associated with the  $n$- and $\bar n$-sectors  we rewrite 
it as a product 
\begin{equation}
O^{(k)}=O^{(k_{1},n)}O^{(k_{2},\bar{n})}.
\end{equation}
Then the  calculation of the $T$-product in Eq.(\ref{Ok-gen}) can be carried out
independently in  each collinear sector considering the soft and collinear
fields as external
\begin{equation}
T\left\{  O^{(k)},\mathcal{L}_{\text{int}}^{(l_{1},n)},\mathcal{L}%
_{\text{int}}^{(m_{1},\bar{n})},...\right\}  =T\left\{  O^{(k_{1}%
,n)},\mathcal{L}_{\text{int}}^{(l_{1},n)},...\right\}  T\left\{
O^{(k_{2},\bar{n})},\mathcal{L}_{\text{int}}^{(m_{1},\bar{n})},...\right\}  .
\label{T=TT}
\end{equation}
Therefore we can perform the analysis of the $T$-products in each sector and
then combine them into the complete soft-collinear operator as in Eq.(\ref{Ok-gen}). The analysis in the
each collinear sector is quite similar because the incoming and outgoing nucleon  states have the same quantum numbers.  Therefore 
 one can consider only the one collinear sector.  To be specific   we  consider the sector associated with the outgoing nucleon.  
The corresponding  $T$-product in Eq.(\ref{T=TT}) must have  the following structure
\begin{equation}
T\left\{  O^{(k_{1},n)},\mathcal{L}_{\text{int}}^{(l_{1},n)},...,\mathcal{L}_{\text{int}}^{(l_{i},n)}\right\}  =O_{n}^{(6)}\ast J_{n}\ast O^{out}_{S} \sim\mathcal{O}(\lambda^{6}), 
\label{TOn}%
\end{equation}
where we suppose  that the leading-order contribution includes  the leading collinear operator $O_{n}^{(6)}$ and has the order  $\mathcal{O}(\lambda^{6})$.  
Such picture is true if  we assume that the   $T$-product (\ref{TOn}) can not have the behavior of order $\lambda^{k}$ with $k<6$. 
Indeed, from an analysis given below we will see that this assumption is correct. 

 The full soft operator $O_{S}$ in Eq.(\ref{Ok-gen}) is given by the product of the soft operators 
 originating in  each collinear sector 
 \be{OS}
 O_{S}=O^{out}_{S}O^{in}_{S}.
 \ee  
where the indices $in$ and $out$ denotes the  appropriate soft part, see Eq. (\ref{TOn}). 
 This operator must  have the nonzero  matrix element  $\langle 0| O_{S}|0\rangle$  in order to describe the  nontrivial  soft-overlap configuration. 

From   Eq.(\ref{TOn}) it follows that the order of the operator $ O^{(k_{1},n)}$  is restricted by values $k_{1}<6$. However one can show that 
it is enough  to consider the operators with $k_{1}<4$.  In order to see this consider the contribution of the operator $ O^{(4,n)}$. 
The number of the different insertions $\mathcal{L}_{\text{int}}^{(l_{i},n)}$ in Eq.(\ref{TOn})   is restricted in general case by the requirement   
\begin{equation}
k_{1}+l_{1}+...+l_{i}\leq 6.
\end{equation}
 Therefore the relevant contribution of  any operator
$O^{(4,n)}$ can be described by the following $T$-products
\begin{equation}
T\left\{  O^{(4,n)},\mathcal{L}_{\text{int}}^{(1,n)}\right\}  ,\ T\left\{  O^{(4,n)},\mathcal{L}_{\text{int}}^{(1,n)},\mathcal{L}_{\text{int}}^{(1,n)}\right\},\  T\left\{  O^{(4,n)},
\mathcal{L}_{\text{int}}^{(2,n)}\right\} . 
 \label{TO4L2L11}
\end{equation}

In order to match  the structure in Eq.(\ref{TOn}) we  always  need to obtain at least  three collinear quarks. 
 It is easy to see that the operator $O^{(4,n)}$ can  include  only  one collinear quark field because  two collinear fields behave as
~$\bar{\xi}_{n}^{c}\bar{\xi}_{n}^{c}\sim\mathcal{O}(\lambda^{4})$. However  the operator $O^{(4,n)}$ must have 
at least  one hard-collinear field.   In order to obtain  the remaining two collinear fields  $\bar{\xi}_{n}^{c}$ one needs  the
insertion of the interaction vertices  generated by the effective Lagrangian. The simplest vertex which
allows this can be generated from the leading order contribution (\ref{Ln0})
$\mathcal{L}_{\xi\xi}^{(0,n)}\rightarrow\mathcal{L}_{\text{int}}^{(1,n)}[\bar \xi^{c}\xi]$ with the help of the
substitution (\ref{sub})  and therefore has at least the dimension one $\mathcal{L}_{\text{int}}^{(1,n)}$. 
Hence we need at least two such  insertions
\be{O4L1cL1c}
T\left\{  O^{(4,n)},\mathcal{L}_{\text{int}}^{(1,n)}[\bar \xi^{c}\xi],\mathcal{L}_{\text{int}}^{(1,n)}[\bar \xi^{c}\xi] \right\}\sim \mathcal{O}(\lambda^{6}),
\ee
which already provides the contribution of order $\lambda^{6}$.  The other two $T$-products in Eq.(\ref{TO4L2L11})  can not provide the required collinear structure. 
However the insertions in Eq.(\ref{O4L1cL1c}) do not include the  soft fields and therefore do not match (\ref{TOn}).\footnote{ One  also  needs 
 the interactions with the soft fields in order to provide the contractions for  all hard-collinear quark fields. } 
  In order to
generate the nontrivial soft operator $O_{S}^{out}$  one needs at least  one insertion with the soft  quark or soft transverse gluon field.
 A  contribution with the soft quark   can be
generated by the insertion of  one of the following vertices  
\begin{equation}
\mathcal{L}_{\text{int}}^{(1,n)}[\bar{\xi}A_{\bot}q ]\simeq\int d^{4}x~\bar{\xi}_{n}\Dsl[A]{5}_{\bot}
q,  \label{xiAq}%
\end{equation}
\begin{equation}
\mathcal{L}_{\text{int}}^{(2,n)}[\bar{\xi}^{c}A_{\bot}q]\simeq\int d^{4}x~\bar{\xi}_{n}^{c}\Dsl[A]{5}_{\bot}q.~~\ \label{xicAq}%
\end{equation}
The  lowest order insertion  with the soft gluon can be obtained from the vertex 
\begin{equation}
\mathcal{L}_{\text{int}}^{(1,n)}[\bar{\xi}A_{\bot} A^{s}_{\bot}\xi]\simeq\int d^{4}x~\bar{\xi}_{n}\Dsl[A]{5}_{\bot}(\bar n\cdot\partial)^{-1}\Dsl[A]{5}_{\bot
}^{s}\xi_{n}.
\end{equation}
where   we use $\xi^{hc}\equiv \xi$  and $A^{hc}\equiv A$  for the hard-collinear quark and gluon fields for simplicity  of notation. 
We also show a simplified structure of the interaction vertices neglecting the Wilson lines.

One can see that inserting  at least  one vertex with the soft field in the $T$-product in Eq.(\ref{O4L1cL1c}) costs  at least  one more factor $\lambda$
 and  will give the subleading result of order  $\lambda^{7}$ or higher.   
 Therefore we conclude that  the operators $O^{(4,n)}$  cannot  provide  the relevant contribution as in Eq.(\ref{TOn}). 

A similar argument allows one to exclude the operator  $O^{(5,n)}$.  The relevant  $\mathcal{O}(\lambda^{6})$ contribution  is given by 
\begin{equation}
T\left\{  O^{(5,n)},\mathcal{L}_{\text{int}}^{(1,n)}\right\}. 
 \label{TO5L1}
\end{equation}
One can easily see that  the required structure (\ref{TOn}) can be described only  by the insertion of the vertex  (\ref{xiAq}) 
 but this gives already a subleading contribution.

In order to perform the analysis of the operators $O^{(k_{1},n)}$ with $k_{1}<4$  we need their explicit expressions. 
The set of the relevant SCET operators  associated with the $n$-collinear
sector can be constructed from the gauge invariant combinations of the
hard-collinear and collinear fields as given in Eqs.(\ref{Chi})-(\ref{AC}). In what follows we do not write for
simplicity  the index $hc$ for the hard-collinear fields assuming $\bar{\chi
}_{n}^{hc}\equiv\bar{\chi}_{n}$ and the same for the gluons. The collinear
fields can be described only by the quark fields $\bar{\chi}_{n}^{c}$ in order
to match the structure (\ref{TOn}). Let us describe the set of  appropriate
operators in the following way
\begin{equation}
O^{(1,n)}=\left\{  \bar{\chi}_{n},~\mathcal{A}_{\bot}^{n}\right\}
\sim\mathcal{O}(\lambda), \label{O1n}
\end{equation}
\begin{equation}
O^{(2,n)}=\left\{  \bar{\chi}_{n}\mathcal{A}_{\bot}^{n},~\bar{\chi}
_{n}\bar{\chi}_{n}\right\}  \sim\mathcal{O}(\lambda^{2}), \label{O2n}
\end{equation}
\begin{equation}
O^{(3,n)}=\left\{  \bar{\chi}_{n}^{c}\bar{\chi}_{n},~\bar{\chi}_{n}
^{c}\mathcal{A}_{\bot}^{n},~\bar{\chi}_{n}\left(  n\cdot\mathcal{A}
^{n}\right)  ,~\bar{\chi}_{n}\mathcal{A}_{\bot}^{n}\mathcal{A}_{\bot}
^{n},~\bar{\chi}_{n}\bar{\chi}_{n}\mathcal{A}_{\bot}^{n},\bar{\chi}_{n}
\bar{\chi}_{n}\bar{\chi}_{n}\right\}  \sim\mathcal{O}(\lambda^{3}).
\label{O3n}
\end{equation}
For simplicity, we do not show the indices and the arguments of the fields.

Let us first consider the operators $O^{(3,n)}$ listed in (\ref{O3n}).   In this
case  the  $T$-products of order $\mathcal{O}(\lambda^{6})$  can be constructed  in the following way
\begin{align}
T\left\{  O^{(3,n)},\mathcal{L}_{\text{int}}^{(3,n)}\right\},\label{O3L3}  \\ 
T\left\{  O^{(3,n)},\mathcal{L}_{\text{int}}^{(2,n)},\mathcal{L}_{\text{int}}^{(1,n)}\right\},\label{O3L2L1} \\ 
T\left\{  O^{(3,n)},\mathcal{L}_{\text{int}}^{(1,n)},\mathcal{L}_{\text{int}}^{(1,n)},\mathcal{L}_{\text{int}}^{(1,n)}\right\}, \label{O3L1L1L1} 
\end{align}
Let us for the moment  postpone the consideration of the  $T$-products which are  less suppressed than $\lambda^{6}$. 
In Eqs.(\ref{O3L3})-(\ref{O3L1L1L1}) we assume that one can also insert the vertices generated by the leading order
Lagrangian $\mathcal{L}^{(0,n)}$ which do not change the scaling behavior.
 It follows  from Eq.(\ref{TOn}) that all these $T$-products must include at
least  one soft field (quark or gluon) and three collinear quark fields.
 For brevity  each operator in the list (\ref{O3n}) is denoted  as
$O_{i}^{(3,n)}$ where the index $i$ corresponds to the place of the operator
in the list (\ref{O3n}) and  $i=1,...,6$. 

The operators $O_{i}^{(3,n)}$ with $i\geq3$ consist only of
hard-collinear fields. Hence in order to produce three collinear quarks one
has to insert at least three vertices $\mathcal{L}_{\text{int}}^{(1,n)}[\bar \xi^{c}\xi]$. This
already yields the contribution of order $\lambda^{6}$ but without the soft spectator
fields.  Therefore  one has to insert at least one time the vertex
$\mathcal{L}_{\text{int}}^{(2,n)}[\bar{\xi}^{c}A_{\bot}q]$ instead of $\mathcal{L}_{\text{int}}^{(1,n)}[\bar \xi^{c}\xi]$ in order
to get a soft spectator. But such substitution yields the contribution of
order $\lambda^{7}$. 

The other  possibility is to insert the vertex
$\mathcal{L}_{\text{int}}^{(3,n)}$ as in Eq.(\ref{O3L3})  which includes  all required  collinear and soft fields.
However one can easily find  that the required  vertex can not have such small order.   This vertex must include 
 $\bar{\xi}_{n}^{c}\bar{\xi}_{n}^{c}\bar{\xi}_{n}^{c}q \dots\sim
\mathcal{O}(\lambda^{9+\dots})$ where dots denote the contribution of the hard-collinear fields. 
Taking into account the scaling of the hard-collinear measure  $d^{4}x\sim\mathcal{O}(\lambda^{-4})$
one obtains that such  vertex is already of order $\mathcal{O}(\lambda^{5+\dots})$.  Hence the operators $O_{i}^{(3,n)}$ with $i\geq3$ can be
omitted. 

 Hence we reduce the list (\ref{O3n}) to the two operators with $i=1,2$ which
include the collinear quark field.
 The $T$-product of these operators with  $\mathcal{L}^{(3,n)}$ as in Eq.(\ref{O3L3})  
can  again be excluded by the same arguments  as before.  The inserted vertex must include  two collinear quark
fields, soft field and at least one hard-collinear field.  Therefore such vertex is of order $\lambda^{4}$ or higher.

The interaction vertices in the  $T$-product in Eq.(\ref{O3L1L1L1}) must
include two collinear fields and at least one soft field. We find the following combination 
\begin{equation}
T\left\{  \bar{\chi}_{n}^{c}\bar{\chi}_{n},
\mathcal{L}_{\text{int}}^{(1,n)}[\bar \xi^{c}\xi],
\mathcal{L}_{\text{int}}^{(1,n)}[\bar \xi^{c}\xi],
\mathcal{L}_{\text{int}}^{(1,n)}[\bar{\xi}A_{\bot}q ] \right\} , 
\label{O31-111q}
\end{equation}
where   the vertices $\mathcal{L}_{\text{int}}^{(1,n)}[\bar \xi^{c}\xi]$ are obtained from the 
leading SCET-I Lagrangian $\mathcal{L}_{\bar{\xi}\xi}^{(0,n)}$ using substitution (\ref{sub}). 
 These  vertices have the following   structure:
\begin{equation}
\mathcal{L}_{\text{int}}^{(1,n)}[\bar \xi^{c}\xi]=\mathcal{L}_{\text{int}}^{(1,n)}[\bar \xi^{c}(A\cdot n)\xi]
=\int dx~\bar{\xi}_{n}^{c}(A_{n}\cdot n)\xi_{n},
\end{equation}
or
\begin{equation}
\mathcal{L}_{\text{int}}^{(1,n)}[\bar \xi^{c}\xi] =\mathcal{L}_{\text{int}}^{(1,n)}[\bar \xi^{c}A_{\bot} A_{\bot}\xi] = \int dx~\bar
{\xi}_{n}^{c}\Dsl[A]{5}_{\bot}(\bar n \cdot \partial)^{-1}\Dsl[A]{5}_{\bot}\xi_{n}.
\label{L1n}
\end{equation}
The structure of $\mathcal{L}_{\text{int}}^{(1,n)}[\bar{\xi}A_{\bot}q ]$ is described in
Eq.(\ref{xiAq}).  The $T$-product in Eq.(\ref{O31-111q}) has always the odd number of the  transverse hard-collinear gluon fields $A_{\bot}$. 
In order to contract them one has to  insert  in Eq.(\ref{O31-111q}) the leading order three gluon vertex 
$\mathcal{L}_{\text{int}}^{(0,n)}\sim \partial_{\bot} A_{\bot}A_{\bot}A_{\bot}$.  Such configuration  provides  higher 
order  loop diagrams.  However the single insertion of the three gluon vertex introduces the  linear
transverse hard-collinear momentum ($\partial_{\bot}\rightarrow p_{hc \bot}$) in the numerator of the loop diagrams 
which can not be  contracted and  therefore such integrals vanish due to the rotational invariance.  
Therefore  the $T$-product in (\ref{O3L1L1L1})  can not
provide the contributions of order $\lambda^{6}$.

We next discuss  possibility  provided by the $T$-product  of Eq.(\ref{O3L2L1}).  The configurations
which matches the structure (\ref{TOn}) read
\begin{equation}
T\left\{  \bar{\chi}_{n}^{c}\bar{\chi}_{n},
\mathcal{L}_{\text{int}}^{(2,n)}[\bar{\xi}^{c}A_{\bot}q],
\mathcal{L}_{\text{int}}^{(1,n)}[\bar \xi^{c}A_{\bot} A_{\bot}\xi]  \right\}, 
\label{TO31-12q}%
\end{equation}
where $\mathcal{L}_{\text{int}}^{(2,n)}[...]$ is given in Eq.(\ref{xicAq}). 
In this case one must again  insert  $\mathcal{L}_{\text{int}}^{(0,n)}\sim \partial_{\bot} A_{\bot}A_{\bot}A_{\bot}$
in order to contract the gluon hard-collinear fields. 
However  the resulting loop  integral is again trivial due to the  rotational invariance as in  the previous case. 

Above consideration also holds  for  the $T$-products which 
have the order $\lambda^{5}$ or smaller because in this case the number of the possible insertions $\mathcal{L}_{\text{int}}$  are
even smaller  and  as a result it is not possible  to obtain the required structure (\ref{TOn}). 

 Therefore we demonstrated  that the operators $O^{(3)}$ in Eq.(\ref{O3n})   provide  only
the subleading contributions of order $\mathcal{O}(\lambda^{7})$ or higher and
therefore can be neglected.

For the  $T$-products with the  operators $O^{(2,n)}$ listed in Eq.(\ref{O2n}) one can consider  at order $\mathcal{O}(\lambda^{6})$   the following 
expressions:
\begin{align}
&  T\left\{  O^{(2,n)},\mathcal{L}_{\text{int}}^{(4,n)}\right\}  ,\label{TO2-L4}\\
&  T\left\{  O^{(2,n)},\mathcal{L}_{\text{int}}^{(3,n)},\mathcal{L}_{\text{int}}^{(1,n)}\right\}
,\label{TO2-L31}\\
&  T\left\{  O^{(2,n)},\mathcal{L}_{\text{int}}^{(2,n)},\mathcal{L}_{\text{int}}^{(2,n)}\right\}  ,
\label{TO2-L22}%
\end{align}%
\begin{equation}
T\left\{  O^{(2,n)},\mathcal{L}_{\text{int}}^{(2,n)},\mathcal{L}_{\text{int}}^{(1,n)},\mathcal{L}_{\text{int}}
^{(1,n)}\right\}  , \label{TO2-L211}%
\end{equation}%
\begin{equation}
T\left\{  O^{(2,n)},\mathcal{L}_{\text{int}}^{(1,n)},\mathcal{L}_{\text{int}}^{(1,n)},\mathcal{L}_{\text{int}}
^{(1,n)},\mathcal{L}_{\text{int}}^{(1,n)}\right\}  . \label{TO2-4xL1}%
\end{equation}
We  again  skip the  $T$-products which are less suppressed  because  they are  stronger restricted.        

In order to obtain the  structure (\ref{TOn})  the vertex
$\mathcal{L}^{(4,n)}$ in Eq.(\ref{TO2-L4}) must include the three collinear quark fields,  one soft fields and two hard-collinear fields because the operators $O^{(2,n)}$ 
are built only from the hard-collinear fields.  However the vertex with such field content is suppressed at least  as $\mathcal{O}(\lambda^{7})$. 
Therefore the $T$-product (\ref{TO2-L4}) can not  provide the  required  structure and can be excluded.

The $T$-product in Eq.(\ref{TO2-L31}) can also be excluded using the similar
counting arguments.  Consider, for instance,  the following $T$-product
\begin{equation}
T\left\{  O^{(2,n)},\mathcal{L}_{\text{int}}^{(3,n)},\mathcal{L}_{\text{int}}^{(1,n)}[\bar\xi^{c}\xi] \right\} .
\end{equation}
In order to obtain the required structure  the interaction  $\mathcal{L}^{(3,n)}_{\text{int}}$ must  include two collinear quark fields,  soft field,
and two hard-collinear fields. Hence such vertex  is at least of order $\mathcal{O}(\lambda^{4})$.  The similar arguments  allows one to exclude
 other similar combinations. Therefore the $T$-product (\ref{TO2-L31}) can be also neglected.

The expression in Eq.(\ref{TO2-L22}) can not describe the structure (\ref{TOn})
because the different interactions $\mathcal{L}^{(2,n)}$ can include only  one collinear
quark field.   Therefore one needs the insertion at least one more vertex with collinear quark field in $T$-product  (\ref{TO2-L22})  that
yields already the higher order contribution.

The analysis of the expressions in (\ref{TO2-L211}) and (\ref{TO2-4xL1}) is
quite similar to the analysis of the analogous operators $O_{1,2}^{(3,n)}$.
 One can consider the operators $O_{1,2}^{(2,n)}$ as the similar operators
which contribute to the same matrix elements at higher orders in the effective theory.
Using the same arguments as in the case of  $O_{1,2}^{(3,n)}$ one finds that in
case of (\ref{TO2-L211}) only the following $T$-product is consistent with the
structure (\ref{TOn})
\begin{equation}
T\left\{  \bar{\chi}_{n}\bar{\chi}_{n},
\mathcal{L}_{\text{int}}^{(2,n)}[\bar{\xi}^{c}A_{\bot}q],
\mathcal{L}_{\text{int}}^{(1,n)}[\bar\xi^{c}\xi],
\mathcal{L}_{\text{int}}^{(1,n)}[\bar\xi^{c}\xi],
\mathcal{L}_{\text{int}}^{(0,n)}\right\} ,
\label{O2-L211L0}
\end{equation}
where $\mathcal{L}_{\text{int}}^{(2,n)}[...]$ is given in Eq.(\ref{xicAq}).  In Eq.(\ref{O2-L211L0}) we 
again added the vertex $\mathcal{L}_{\text{int}}^{(0,n)}\sim \partial_{\bot} A_{\bot}A_{\bot}A_{\bot}$
  in order to contract  the  hard-collinear gluon fields $A_{\perp}^{n}$. 
 However the  this vertex generate the hard-collinear transverse
 momentum in the numerator and the corresponding integrals vanish due to the
rotation invariance. The same conclusion is also valid  for the $T$-product
(\ref{TO2-4xL1}). Hence the operators $O_{1,2}^{(3,n)}$ in Eq.(\ref{O2n})  can not provide the contributions of order $\mathcal{O}(\lambda^{6})$  and can be neglected.
This is also true for the $T$-products  with the smaller number of insertions  $\mathcal{L}_{\text{int}}$ which have the order  $\mathcal{O}(\lambda^{p})$ with $p<6$. 

We reduced the  set of the SCET operators describing the
soft-overlap contribution to the operators $O^{(1,n)}$ in Eq.(\ref{O1n}).
Let us consider the quark operator.  
For our purpose it is enough to demonstrate that  there is at least one $T$-product 
which provides the soft-collinear operator as in Eq.(\ref{TOn}).  
We suggest to consider the following    $T$-product   
\begin{equation}
T\{\bar{\chi}_{n},
\mathcal{L}_{\text{int}}^{(2,n)}[\bar{\xi}^{c}A_{\bot}q],
\mathcal{L}_{\text{int}}^{(2,n)}[\bar{\xi}^{c}A_{\bot}(\bar n\cdot A)q],
\mathcal{L}_{\text{int}}^{(1,n)}[\bar \xi^{c}A_{\bot} A_{\bot}\xi],
 \label{TO1q}%
\end{equation}
where $\mathcal{L}_{\text{int}}^{(2,n)}[\dots]$ is given in Eq.(\ref{xicAq}), $\mathcal{L}_{\text{int}}^{(1,n)}[\bar \xi^{c}A_{\bot} A_{\bot}\xi]$ in Eq.(\ref{L1n}). 
 The  hard-collinear gluon field $(\bar n\cdot A)$  
in the argument of $\mathcal{L}_{\text{int}}^{(2,n)}[\dots]$ appears  from the hard-collinear Wilson line.  
Notice that  configuration described in Eq(\ref{TO1q})  includes the two soft quark fields and therefore can
be associated with the two soft spectators.  The complete operator describing
the soft-overlap contribution in Eq.(\ref{TJJscet}) can be constructed from
the two hard-collinear operators as
\begin{equation}
O^{\sigma}=\bar{\chi}_{n}\gamma^{\sigma}_{\bot}\chi_{\bar{n}},
\end{equation}
where we took into account the helicity conservation in the hard subprocess.  
 A more detailed study of this soft-spectator contribution  was  carried out in
Refs.\cite{Kivel:2010ns,Kivel:2012mf} .

The contribution of the  gluon operator $\mathcal{A}_{\bot}^{n}$ is
suppressed. In order to describe the structure  (\ref{TOn}) one has to
convert the hard-collinear gluon into hard-collinear or collinear quarks. Such
a conversion can be done by insertion of the vertex with the soft quark field
$\sim\bar{\xi}_{n}A_{\perp}q$  that  provides  the extra factor $\lambda$
comparing to the hard-collinear quark operator $\bar{\chi}_{n}$. Nevertheless
taking into account that this suppression is associated only with the
hard-collinear dynamics one can expect that the gluon matrix element can
provide a sizeble numerical effect in the region where the hard-collinear
scale is not large. However the complete gluon operator has two transverse indices and therefore the 
 corresponding matrix element can be  parametrized  only in terms of the chiral-odd combinations:
\begin{equation}
\left\langle p^{\prime}\right\vert \mathcal{A}_{\bot\alpha}^{n}\mathcal{A}%
_{\bot\beta}^{\bar{n}}\left\vert p\right\rangle _{\text{{\footnotesize SCET}}%
}=g_{\alpha\beta}^{\bot}~\bar{N}_{n}N_{\bar{n}}~\mathcal{F}^{g}(t)+\epsilon
_{\alpha\beta}^{\bot}~\bar{N}_{n}\gamma_{5}N_{\bar{n}}~\mathcal{\tilde{F}%
}^{g}(t).
\end{equation}
Probably  this operators can contribute only to the helicity flip amplitudes  which we do not consider in the present 
publication.

Therefore we demonstrated that the soft-overlap contribution arising in the
$T$-product of the two electromagnetic current and overlapping with the
leading power hard-spectator contribution is described by the single SCET
operator constructed from  two quark fields $O^{\sigma}=\bar{\chi}
_{n}\gamma_{\bot}^{\sigma}\bar{\chi}_{n}$.

There are also  soft-overlap contributions involving the photon states.   
However these contributions  can not describe the soft-overlap contribution which can
mix with the hard hard-spectator interaction in Eq.(\ref{TJJscet}). 
 The consideration of the corresponding  SCET operators  are presented in  Appendix~A.  
 We obtained  that these configurations are also power suppressed.

\section{Calculation of the hard coefficient functions for the soft spectator contribution}
\label{calculation}

In order to compute the coefficient functions $C_{i}$ in Eq. (\ref{Ai-fact})  we consider the auxiliary process $\gamma q\rightarrow \gamma q$ in the wide angle kinematics.  Taking the   matrix elements in  Eq.(\ref{TJJscet})  gives 
\begin{equation}
\int dx~e^{i(q^{\prime}x)}\left\langle q(p^{\prime})\right\vert
T\left\{  J^{\mu}(x)J^{\nu}(0)\right\}  \left\vert q(p)\right\rangle
=C^{\mu\nu\sigma}(s,t) \left\langle q(p^{\prime})\right\vert O^{\sigma}\left\vert q(p)\right\rangle _{\text{\tiny SCET}},
\label{TJJme}  
\end{equation}
where the quark momenta $p, \ p'$ are given by the massless approximation in Eq.(\ref{mom1}). 
The hard coefficient function $C^{\mu\nu\sigma}$  can be presented  as  a series in the strong coupling $\alpha_{s}$
\be{Cmns}
C^{\mu\nu\sigma} (s,t)=C_{\text{\tiny LO}}^{\mu\nu\sigma}(s,t)+\frac{\alpha_{s}}{\pi }C_{\text{\tiny NLO}}^{\mu\nu\sigma}(s,t)+\mathcal{O}(\alpha^{2}_{s}).
\ee  
The scalar  coefficients  $C_{i}$ in Eq.(\ref{Ai-fact})  can be computed from $C^{\mu\nu\sigma}$  with the help of Eqs.(\ref{TiTj})-(\ref{TiTi}). 
The hard coefficient functions $C_{i}$ are  functions of the Mandelstam variables describing the quark scattering.  They can also be represented 
 as functions of  the scattering angle $\theta$  and energy $s$.  
Below  we present  the expressions for $C_{i}$  in terms of the variables $s, t$ and $u$  assuming 
 that these are massless or partonic variables which  satisfy  the massless  relations
\begin{equation}
t=-\frac{s}{2}(1-\cos\theta),~u=-\frac{s}{2}(1+\cos\theta),~s+u+t=0.
\label{stu0}
\end{equation}  
For simplicity we will not introduce for them special partonic notations assuming that this feature is clear  and does not lead to any confusion.  

\begin{figure}[ptb]%
\begin{center}
\includegraphics[width=4.5in]{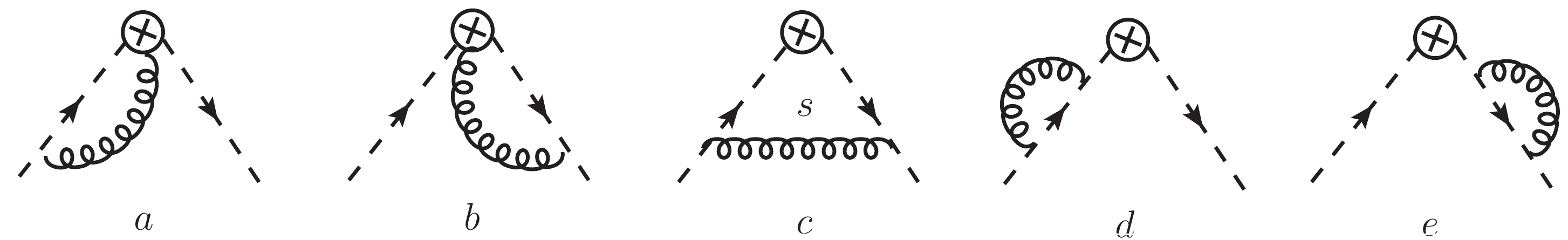}
\end{center}
\caption{The next-to-leading SCET diagrams describing the contribution
$\hat{\mathcal{F}}_{\text{NLO}}^{q}$ in SCET-I.} The vertex of the operator is shown by the crossed circle.  The dashed lines denote the hard-collinear quarks,  the  gluon lines in diagrams $(a,b,d,e)$ are also hard-collinear.  The soft gluon line in diagram   $(c)$ is indicated by index $s$. 
\label{nlo-scet}%
\end{figure}

The  matrix element in the {\it rhs} of Eq.(\ref{TJJme})  can be parametrized as 
\be{F1q}
\left\langle q(p^{\prime})\right\vert O^{\sigma}\left\vert q(p)\right\rangle _{\text{\tiny SCET}} =\bar u(p')\nbn  \gamma^{\sigma}_{\perp}\nbn u(p)\ e^{2}_{q} \ \hat{\mathcal{F} }^{q}(t)
\equiv \bar u_{n}\gamma^{\sigma}_{\perp} u_{\bar n}\ e^{2}_{q}\ \hat{\mathcal{F} }^{q}(t),
\ee
where $u(p)$,  $\bar u(p') $ denote the quark wave functions, $e^{2}_{q}$ is the charge of the quark (we consider  one flavor for simplicity).  
The quark SCET  FF  $\hat{\mathcal{F} }^{q}$ can be computed in SCET-I  order by order in  perturbation theory. Therefore it can be presented as 
\be{F1inPT}
\hat{\mathcal{F} }^{q}(t)=\hat{\mathcal{F} }^{q}_{LO}(t)+\frac{\alpha_{s}}{\pi}\hat{\mathcal{F} }^{q}_{NLO}(t)+\mathcal{O}(\alpha^{2}_{s}).
\ee
The leading-order contribution is given by the tree level vertex diagram and  reads
\be{F1qLO}
\hat{\mathcal{F} }^{q}_{LO}=1.
\ee
   In order to  compute the next-to-leading contribution in Eq.(\ref{F1inPT})  one has to consider  the one-loop diagrams shown in Fig.\ref{nlo-scet}.  

The QCD perturbative expansion on the {\it lhs} of Eq.(\ref{TJJme})  is given by the QCD diagrams describing the Compton scattering on a quark $\gamma q\rightarrow \gamma q$.   The leading order contribution is given by the tree  diagrams with massless quarks.  Computing these diagrams and comparing with the  {\it rhs} of Eq.(\ref{TJJme}) one obtains the  expression for the leading-order coefficient functions 
\begin{align}
C_{2}^{\text{{\tiny LO}}}(s,t) &  =-C_{4}^{\text{{\tiny LO}}%
}(s,t)=\frac{s-u}{su}=-\frac{1}{s}\frac{1+\cos^{2}\theta/2}{\sin^{2}\theta
/2},
\label{C24LO} \\
C_{6}^{\text{{\tiny LO}}}(s,t) &  =\frac{t}{su}=\frac{1}{s}\frac
{\cos^{2}\theta/2}{\sin^{2}\theta/2}.
\label{C6LO}
\end{align} 

\begin{figure}[ptb]
\centering
\includegraphics[width=5.472in]
{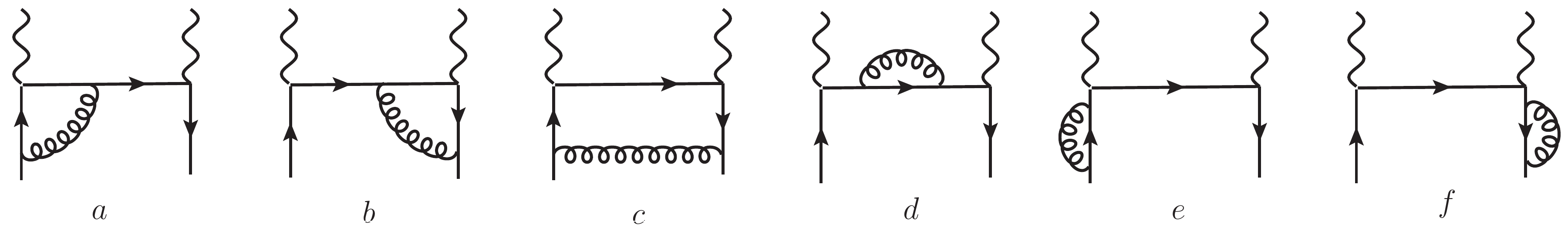}%
\caption{Next-to-leading order QCD diagrams required for the calculation of the $C^{\mu\nu\sigma}_{NLO}$. The crossed diagrams are not shown for simplicity. }
\label{nlo-digrams}%
\end{figure}

In order to obtain the $C_{\text{\tiny NLO}}^{\mu\nu\sigma}$ in Eq.(\ref{Cmns}) one has to compute the one-loop corrections to the  matrix elements in Eq.(\ref{TJJme}).   
The QCD one-loop diagrams for the $T$-product of the electromagnetic currents  are shown in Fig.\ref{nlo-digrams}.  

 As  we  already explained the leading order tree diagrams can be considered  with the on-shell  massless quarks. 
However beyond the tree level  the situation is more complicated due to the IR- and UV-divergencies   which must be regularized.  The most convenient technique  is to use the dimensional regularization (DR) with $D=4-2\varepsilon$ in order to treat   all  singularities which appear in the diagrams.  We will use  DR and the $\overline{\text{MS}}$-scheme in order to perform UV-renormalization of the matrix element $\langle O^{\sigma}\rangle $ on {\it rhs} of Eq.(\ref{TJJme}).  The sum of the QCD diagrams in Fig.\ref{nlo-digrams} is UV-finite because of the non-renormalization theorem for the electromagnetic current. However  UV-divergencies are presented in the individual  vertex ($a,b$) and self-energy ($d,e,f$) diagrams in Fig.\ref{nlo-digrams} .  
We also use DR in order to evaluate  corresponding  UV-divergent integrals.  

In order to treat IR-divergencies in the  QCD diagrams in Fig.\ref{nlo-digrams} we  consider the off-shell momenta for  the outgoing quarks. The corresponding  IR-singularities are 
 logarithmic and  therefore we  consider  the small off-shell momenta only in the denominators of the propagators  and use  the on-shell momenta in the numerators of the integrands. 
The nice feature of such approach is that  UV-divergent  integrals are IR-finite (because corresponding integrals are logarithmic) 
and they can be easily singled out. Therefore  the IR-divergent contributions are UV-finite  and  can be computed in $D=4$ which allows one to easily perform the required manipulations with the Dirac algebra.  Following this way  we  avoid a discussion  about  consistent representation of the matrix $\gamma_{5}$ in $D$ dimensions  and possible  mixing with the so-called evanescent operators.  On the other hand  such IR-regularization introduces  more complicated integrals giving the  IR-logarithms. 

The sum of the  QCD diagrams in Fig.\ref{nlo-digrams}  can be interpreted as radiative corrections to the
hard coefficient function and to the SCET-I matrix element
\begin{equation}
\sum_{i} D_{i}^{\mu\nu}=C_{\text{{\tiny NLO}}}^{\mu\nu\sigma}\ 
(s,t;\mu_{F}^{2})~ \bar u_{n}\gamma^{\sigma}_{\perp} u_{\bar n}\ e^{2}_{q}\hat{\mathcal{F} }_{LO}^{q}+C_{\text{{\tiny LO}}}^{\mu\nu\sigma
}(s,t)~\bar u_{n}\gamma^{\sigma}_{\perp} u_{\bar n}\ e^{2}_{q}\hat{\mathcal{F} }_{NLO}^{q}(t;\mu_{F}^{2},p^{2},p'^{2}).
\label{sumDi} %
\end{equation}
Here $D_{i}^{\mu\nu}$ denotes the contribution of each diagram in Fig.\ref{nlo-digrams} (up to the simple factor $\alpha_{s}/\pi$).
   In Eq.(\ref{sumDi}) we show explicitly  the dependence on the factorization scale $\mu_{F}$  
and on the IR-regulators $p^{2}$ and $p'^{2}$. 
In order  to obtain $C_{\text{{\tiny NLO}}}^{\mu\nu\sigma}$   we need  to compute the following difference
\begin{equation}
C_{\text{{\tiny NLO}}}^{\mu\nu\sigma}(s,t;\mu_{F}^{2})\ e^{2}_{q}\bar{\xi}_{n}\gamma_{\bot}^{\sigma}\xi_{\bar{n}}  =\sum_{i}
D_{i}^{\mu\nu}
-
\bar u_{n}\gamma^{\sigma}_{\perp} u_{\bar n}\ C_{\text{{\tiny LO}}}^{\mu\nu\sigma}(s,t)\ e^{2}_{q}  \hat{\mathcal{F} }_{NLO}^{q}(t;\mu_{F}^{2},p^{2},p'^{2}). 
\label{Cnlodif}%
\end{equation}
The sum of the QCD diagrams $D_{i}^{\mu\nu}$ depends on  the soft  scales  $p^{2}$ and $p'^{2}$ but does not depend on the factorization scale $\mu_{F}$. 
However the soft scales  must cancel on the {\it rhs} of Eq.(\ref{Cnlodif}).  This cancellation provides  a powerful check of the derived factorization theorem.  
It is clear that the dependence on  the factorization scale  of the 
$C_{\text{{\footnotesize NLO}}}^{\mu\nu\sigma}$ is  related with the  renormalization properties of the  SCET-I operator $O^{\sigma}$.  
The renormalization of  $O^{\sigma}$  was already studied in the literature.  A more detailed discussion  can be found, for instance,  in  Ref.\cite{Manohar:2003vb}.   
This operator  is multiplicatively renormalizable 
and using the independence of the physical amplitude from the factorization scale  
$\mu_{F}\equiv \mu$  one can  derive the RG-equation for the coefficient function 
\be{RGC}
\mu \frac{d}{d\mu}C^{\mu\nu\sigma}(s,t;\mu^{2})=\frac{\alpha_{s}}{4\pi}C_{F} \left\{ \ 4\ln[-t/\mu^{2}] -6   +\mathcal{O}(\alpha_{s})  \right\}C^{\mu\nu\sigma}(s,t;\mu^{2}),
\label{RGC}
\ee  
where as usually $C_{F}=(N^{2}_{c}-1)/2N_{c}$.    The expression in the  figured brackets on the {\it rhs} of Eq.(\ref{RGC}) is the leading-order anomalous dimension of the operator 
 $O^{\sigma}$  \cite{Manohar:2003vb}. 

A  detailed discussion  of   the calculation of the  {\it rhs} in Eq.(\ref{Cnlodif}) is presented in the Appendix~B.  
Here we provide only the main results for the scalar coefficient functions
$C_{i}$ defined in Eq.(\ref{Ai-fact}).   Let us   define their perturbative expansion as  
\begin{equation}
C_{i}(s,t;\mu^{2})=C_{i}^{\text{{\tiny LO}}}(s,t)
+\frac{\alpha_{s}}{4\pi}C_{F}
~C_{i}^{\text{{\tiny NLO}}}(s,t;\mu^{2})+\mathcal{O}(\alpha_{s}^{2}).
\label{CiNLO}
\end{equation}
The leading-order coefficients $C_{i}^{\text{{\tiny LO}}}$ are presented  in Eqs.(\ref{C24LO})-(\ref{C6LO}).  
At the next-to-leading order  we obtained the following expressions
\begin{align}
C_{2}^{\text{{\tiny NLO}}} (s,t;\mu^{2}) &  =C_{2}^{\text{{\tiny LO}}}(s,t) \left\{  -\ln^{2}[-t/{\mu}^{2}]+3\ln
[-t/{\mu}^{2}]\right\}  +\frac{t^{2}-su}{s^{2}u^{2}}\left(  u\ln
^{2}\left[  \frac{u}{t}\right]  -s\ln^{2}\left[  \frac{s}{t}\right]  \right)
\nonumber \\
&  -\left(  \frac{3}{s}+\frac{2}{u}\right)  \ln\left[  \frac{s}{t}\right]
+\left(  \frac{3}{u}+\frac{2}{s}\right)  \ln\left[  \frac{u}{t}\right]
-5\frac{s-u}{su}-\pi^{2}\frac{s-u}{su}\left(  \frac{t^{2}}{su}-\frac{7}
{6}\right)  ,
\label{C2NLO}
\end{align}
\begin{align}
C_{4}^{\text{\tiny NLO}}(s,t;\mu^{2})   &  =C_{4}^{\text{{\tiny LO}}}\left\{-\ln^{2}[-t/{\mu}^{2}]+3\ln[-t/{\mu}^{2}]\right\} 
 -\frac{t^{2}-su}{s^{2}u^{2}}\left(  u\ln^{2}\left[
\frac{u}{t}\right] - s\ln^{2}\left[  \frac{s}{t}\right] \right)
\nonumber  \\
&  +\left(  \frac{3}{s}+\frac{2}{u}\right)  \ln\left[  \frac{s}{t}\right]
-\left(  \frac{3}{u}+\frac{2}{s}\right)  \ln\left[  \frac{u}{t}\right]
+9\frac{s-u}{su}+\pi^{2}\frac{s-u}{su}\left(  \frac{t^{2}}{su}-\frac{7}{6}\right)  ,
\label{C4NLO}
\end{align}
\begin{align}
C_{6}^{\text{\tiny NLO}}(s,t;\mu^{2})  &  =C_{6}^{\text{{\tiny LO}}}
\left\{-\ln^{2}[-t/{\mu}^{2}]+3\ln[-t/{\mu}^{2}]\right\} 
 -\frac{t^{2}+su}{s^{2}u^{2}}\left(  s\ln^{2}\left[  \frac{s}{t}\right] 
  +u\ln^{2}\left[\frac{u}{t}\right]  \right)
\nonumber  \\
&  -\left(  \frac{3}{s}+\frac{2}{u}\right)  \ln\left[  \frac{s}{t}\right]
-\left(  \frac{3}{u}+\frac{2}{s}\right)  \ln\left[  \frac{u}{t}\right]
-7\frac{t}{su}+\pi^{2}\frac{t}{su}\left(  \frac{t^{2}}{su}+\frac{7}{6}\right) .
\label{C6NLO}
\end{align}
One sees that the soft scales $p^{2}$ and $p'^{2}$  are not present in Eqs.(\ref{C2NLO})-(\ref{C6NLO}) because they cancel  as it is required by the factorization. 
One can also easily check that  the scale  dependence in the obtained $C_{i}^{\text{\tiny NLO}}$  is in agreement 
 with the  RG-equation (\ref{RGC}).   The leading-order coefficients  $C_{i}^{\text{\tiny LO}}$ are real but the next-to-leading expressions in Eqs.(\ref{C2NLO}-\ref{C6NLO}) have the 
 nontrivial imaginary part  which appears due to the logarithm $\ln[s/t]\equiv \ln[-s/|t| -i0]$.  
 One can also observe that the expressions for the coefficients $C_{2,4}^{\text{\tiny NLO}}$  differ from their counterparts in the leading-order approximation  not only by the relative sign,   
  but also  by the simple rational  term
 \be{C24NLO}
 C_{2}^{\text{\tiny NLO}}+C_{4}^{\text{\tiny NLO}}=4\frac{s-u}{su}.
 \ee

\section{Phenomenology}
\label{phenomenology}
In order to apply  the factorization formula Eq.(\ref{Ai-fact})  in a  phenomenological analysis  one must  define the unknown nonperturbative form factor $\mathcal{F}_{1}$.
But there is one more difficulty hidden in the factorization expression (\ref{Ai-fact}). This problem   is related to the careful separation of the  soft and collinear region and in developing a 
technique for a systematic resummation of the large rapidity logarithms.  
Let us recall  that the FF $\mathcal{F}_{1}$ implicitly depends on the specific rapidity regularization which helps to separate the soft and 
collinear regions.  At present such a full description  still remains a challenge.  Meanwhile a useful  phenomenological consideration  can be carried out.  
 This is possible  due to  the universality of the definition of the  SCET FF $\mathcal{F}_{1}$ in the factorization approach and due to  its specific properties. 
 This one  unknown quantity describes the three independent amplitudes $T_{2,4,6}$.  Therefore  the idea is to re-express this quantity in terms of  any one amplitude  
 and then use this expression  for the remaining   two  amplitudes.  This allows one to establish the relation between the three amplitudes up to well defined 
 hard-spectator corrections.   

In order to be specific  let us  use  Eq.(\ref{Ai-fact}) and write for  $\mathcal{F}_{1}$ the following expression
\begin{equation}
\mathcal{F}_{1}(t)=\mathcal {R}(s,t)- \mathbf{\Psi}\ast
H_{2}(s,t)\ast \mathbf{\Psi}/C_{2}(s,t).
\label{F1phys}
\end{equation}
Here we define the ratio \begin{equation}
\mathcal{R}(s,t)=\frac{T_{2}(s,t)}{C_{2}(s,t,{\mu}^{2}=-t)}.
\label{Rdef}%
\end{equation}
Note that the {\it rhs} of Eq.(\ref{F1phys}) does not depend on the total energy $s$. 
Substituting this equation in the expressions for the amplitudes $T_{4,6}$ (\ref{Ai-fact}) we obtain
\begin{equation}
T_{i}(s^{\prime},t)=C_{i}(s^{\prime},t) \mathcal{R}(s,t) %
+\mathbf{\Psi}\ast\left\{  H_{i}(s^{\prime},t)-C_{i}(s^{\prime},t)\frac
{H_{2}(s,t)}{C_{2}(s,t)}\right\}  \ast\ \mathbf{\Psi.}
\label{T24R}
\end{equation}
On the left side of this equation we have a well defined physical amplitude therefore the right side must also be well defined. This means that the potential end-point 
singularities in the hard-spectator corrections must cancel in the difference on the {\it rhs} of Eq.(\ref{T24R}). We assume  that all  the hard coefficient functions in  Eq.(\ref{T24R})
are defined at ${\mu}^{2}=-t$.  We also used the different values of the total energy $s'$ in Eq.(\ref{T24R}) in 
order to stress that the substitution  (\ref{F1phys}) does not depend on the energy $s$.   
 
If the values of the  hard-spectator contributions in Eq.(\ref{T24R})  are small, then such terms can be neglected and we obtain  
\begin{equation}
T_{i}(s^{\prime},t)\simeq C_{i}(s^{\prime},t)  \mathcal{R}(s,t)   \Leftrightarrow\frac{T_{i}(s^{\prime},t)}{C_{i}(s^{\prime},t)}\simeq \mathcal{R}(s,t).
\label{Ti-ssp}
\end{equation}
Notice that this formula is valid to all orders in $\alpha_{s}$ for the
coefficient $C_{i}$ but at order $\alpha_{s}^{2}$ one has to take into account the hard-spectator corrections. 
 
 Obviously the choice of the amplitude $T_{i}$  for the definition of the  ratio $ \mathcal{R}$ in (\ref{Rdef}) does not play any essential role.  
 In the definition (\ref{Rdef})  we also used a freedom to  fix the factorization scale and chose $\mu^{2}=-t$ as  simplest realization.    

 If the bulk contribution to the amplitudes $T_{i}(s,t)$ is provided by the
soft-overlap term then one expects that the ratio $\mathcal{R}$ depends only very weakly on
the energy $s$%
\begin{equation}
\frac{d}{ds}\mathcal{R}(s,t)\simeq\mathcal{O}(\alpha_{s}^{2}),\label{dR=0}%
\end{equation}
where the higher order corrections  $\sim\mathcal{O}(\alpha_{s}^{2})$ are again given 
by the hard-spectator contributions and assumed to be  relatively  small.
This  picture can be verified  when comparing with the data. 

Consider the formula   (\ref{dsig})  for the cross section.  Using expressions (\ref{Ti-ssp}) for the amplitudes $T_{2,4,6}$ and neglecting the helicity flip amplitudes 
$T_{1,3,5}\simeq 0$  and power suppressed terms $\sim m/Q$  we obtain 
\begin{equation}
\ \frac{d\sigma}{dt}=\frac{\pi\alpha^{2}}{s^{2}}~|\mathcal{R}(s,t)|^{2}%
(-su)\left(  \frac{1}{2}|C_{2}(s,t)|^{2}+\frac{1}{2}|C_{4}(s,t)|^{2}%
+|C_{6}(s,t)|^{2}\right)  . \label{dsdt}%
\end{equation}
 If the ratio $\mathcal{R}(s,t)$ depends mostly  on the momentum transfer $t$  then  the energy dependence of the cross section in
Eq.(\ref{dsdt}) is defined only  by the hard coefficient functions. To the leading-order
accuracy  using Eqs.(\ref{C24LO}),(\ref{C6LO}) one obtains
\begin{equation}
\ \frac{d\sigma}{dt}\simeq \left. \frac{2\pi\alpha^{2}}{s^{2}}~|\mathcal{R}%
(s,t)|^{2}\left(  \frac{s}{-u}+\frac{-u}{s}\right)\right|_{m=0}  = \frac{d \sigma_{0}^{\text{KN} }}{dt} |\mathcal{R}(s,t)|^{2}.
\end{equation}
Here  only the the ratio $\mathcal{R}$ provides the difference from the point-like Klein-Nishina cross section $d\sigma_{0}^{\text{KN}}$  due to  the  nucleon
structure. This simple leading-order formula is  modified by  the corrections
from the QCD hard subprocess. To the next-to-leading accuracy   one  finds
\begin{equation}
\ \frac{d\sigma}{dt}\simeq  \frac{d\sigma_{0}^{\text{KN}}}{dt}   |\mathcal{R}(s,t)|^{2}
\left(  1+\frac{\alpha_{s}}{4\pi}C_{F}\frac{C_{2}^{\text{\tiny LO} }
\text{Re}\left[  C_{2}^{\text{\tiny NLO}}-C_{4}^{\text{\tiny NLO}}\right]  +C_{6}^{\text{\tiny LO}}\text{Re}\left[
C_{6}^{\text{\tiny NLO}}\right]  }{|C_{2}^{\text{\tiny LO}}|^{2}+|C_{6}^{\text{\tiny LO}}|^{2}}\right)  ,
\label{dsNLO}
\end{equation}
where we used that the leading order coefficient functions are real and
$\left\vert C_{2}^{\text{\tiny LO}}\right\vert =\left\vert C_{4}^{\text{\tiny LO}}\right\vert $.  

This result allows one to extract the absolute value $|R(s,t)|$ and check
the equation  Eq.(\ref{dR=0}) using the experimental data for the cross
section.  Using Eq.(\ref{dsNLO})  one easily obtains  
\begin{equation}
|\mathcal{R}(s,t)|\approx\sqrt{\frac{d\sigma^{\text{exp}}/dt}{d\sigma_{0}^{\text{KN}
} /dt}}\left(  1-\frac{1}{2}\frac{\alpha_{s}}{4\pi}C_{F}\frac{C_{2}^{\text{\tiny LO}
}\text{Re}\left[  C_{2}^{\text{\tiny NLO}}-C_{4}^{\text{\tiny NLO}}\right]  +C_{6}
^{\text{\tiny LO}}\text{Re}\left[  C_{6}^{\text{\tiny NLO}}\right]  }{|C_{2}^{\text{\tiny LO}}
|^{2}+|C_{6}^{\text{\tiny LO}}|^{2}}\right)  .
\label{Rfitm0}
\end{equation}

In obtaining this formula we neglected by the all power corrections  suppressed as $\mathcal{O}(1/Q)$.  Nevertheless the  mass of the nucleon is not too small
as compared to the values of the energy $s$ and momentum transfer $-t$ and  therefore the power corrections may still provide a sizeble numerical  effect.  
In order to estimate their size we  suggest to include in our considerations at least  the so-called  kinematical power corrections.  We define them in the following way.  
We again  neglect  the helicity flip amplitudes $T_{1,3,5}\simeq 0$ but use the exact kinematical coefficients with the nucleon mass $m$ in Eq.(\ref{dsig}).  
In this case we obtain
\bea
\frac{d\sigma}{dt}&=&\frac{\pi\alpha^{2}}{(s-m^{2})^{2}}|\mathcal{R}|^{2}
\left\{
(s-m^{2})(m^{2}-u)\frac{1}{2} (|\bar C_{2}|^{2}+|\bar C_{4}|^{2})+(m^{4}-su)|\bar C_{6}|^{2}
\right\}.
\label{dsPC}
\eea
  The coefficient functions $\bar C_{i}$ in  the expression (\ref{dsPC}) are the functions of $s$ and $t$ which are defined in the exact kinematics with $m\neq 0$ and therefore they include the powers of  $m/s$. We define these functions from the massless expressions $C_{i}(s,\cos\theta)$ computed above in the following way:   
    \be{cosex}
 \bar C_{i}(s, t)=C_{i}\left(s,\cos\theta=1+\frac{2t s}{(s-m^{2})^{2}}\right )=\left. C_{i}(s,\cos\theta)\right|_{m=0}+\mathcal{O}(m/s).
  \ee
 In other words,  in  the massless kinematics in the expressions for  $\left. C_{i}( s, \theta)\right|_{m=0}$  we identify the massless variables with the exact  energy $s$ and scattering angle $\theta$ substituting $\cos\theta=1+\frac{2t s}{(s-m^{2})^{2}}$.  In what follow we denote  the absolute value $|\mathcal{R}|$ which is obtained using  Eq.(\ref{dsPC})  as  $|\bar{\mathcal{R}}|$
 \be{barR}
 |\bar{\mathcal{R}}|=
 \sqrt{
 \frac{d\sigma^{\text{exp}} }{dt}
 }:
 \sqrt{
 \frac{ \pi\alpha^{2} }  { (s-m^{2})^{2} } \left( 
(s-m^{2})(m^{2}-u)\frac{1}{2} (|\bar C_{2}|^{2}+|\bar C_{4}|^{2})+(m^{4}-su)|\bar C_{6}|^{2}
\right)
} .
 \ee
 In the numerical calculations we expanded the  {\it rhs } of  Eq.(\ref{barR}) with respect to $\alpha_{s}$.

The data points for the cross section  $d\sigma^{\text{exp}}$  at large $s$, $-t$ and $-u$ were published  in Ref.\cite{Danagoulian:2007gs}. 
In our analysis we use only  the data for which  $|u|,|t|\geq  2.5$~GeV$^{2}$. 
In computing the next-to-leading contribution we use the running
coupling $\alpha_{s}(\mu^{2}=1.5$~GeV$^{2})=0.360$ with $n_{f}=4$  and define the scale for the
running coupling as $\mu^{2}=\min\{-t,-u\}$.

\begin{figure}[h]
\begin{center}%
{\includegraphics[width=3.5214in]{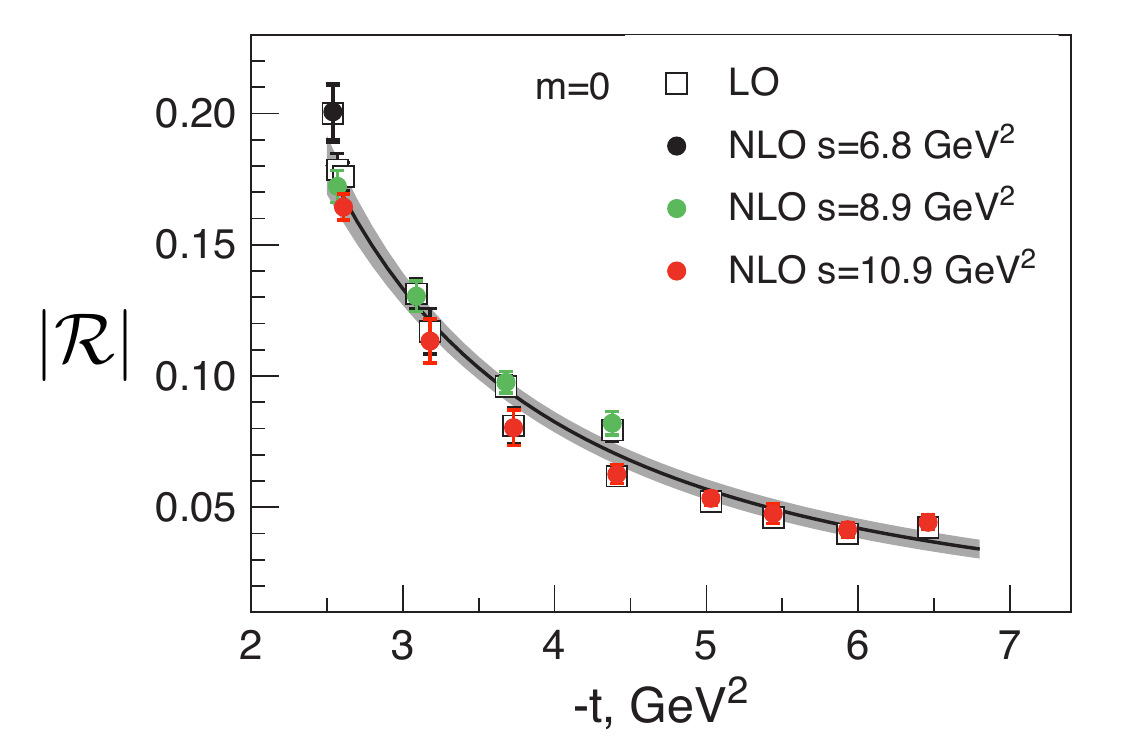}} 
\end{center}
\caption{ The extracted values of $|\mathcal{R}|$ as a functions of the momentum
transfer in the leading-order (open squares) and next-to-leading order (color circles)
approximations for the hard coefficient functions $C_{i}$ .  The solid line demonstrates the fit of
the NLO $|\mathcal{R}|$ using the power behavior as  in Eq.(\ref{Rfit}). The shaded area
shows the $99\%$ confidence bands. }%
\label{Rnlo}%
\end{figure}

\begin{figure}[h]
\begin{center}%
\begin{tabular}[c]{ll}%
{\includegraphics[width=3.3in]{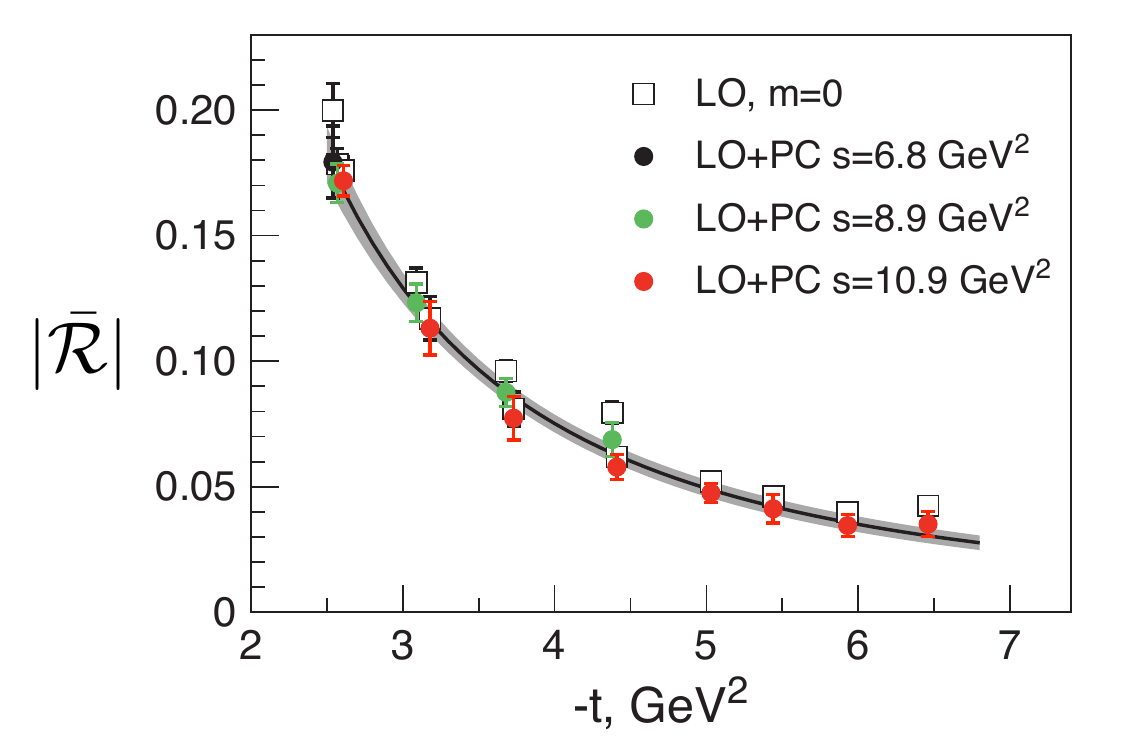}} 
{\includegraphics[width=2.8in]{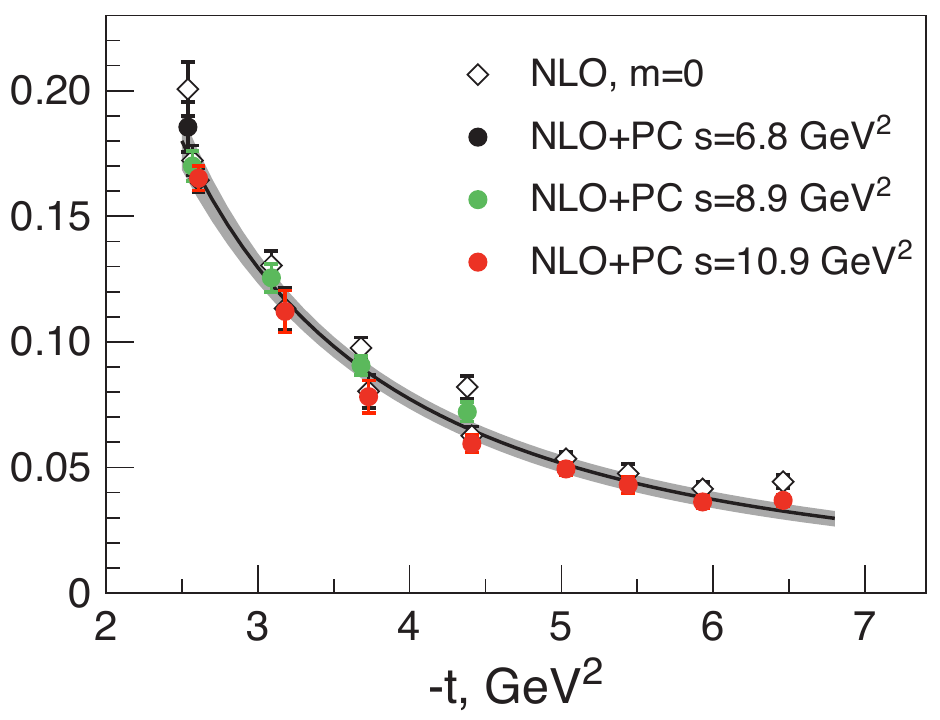}} 
\end{tabular}
\end{center}
\caption{ The extracted values of $|\bar{\mathcal{R}}|$ as a function of the momentum
transfer obtained using the leading-order (left plot) and next-to-leading (right plot) order
approximations for the hard coefficient functions $\bar{C_{i}}$. The open squares (rhombs) show the LO (NLO)  values extracted  with  $m=0$ 
using Eq.(\ref{Rfitm0}).   The solid lines show the fit of  the $|\bar{\mathcal{R}}|$  with the formula in Eq.(\ref{Rfit}). The shaded area
shows the $99\%$ confidence bands. }%
\label{RNLOPC}%
\end{figure}

Our numerical results for the $|\mathcal{R}|$ and  $|\bar{\mathcal{R}}|$ are shown in Fig.\ref{Rnlo}  and Fig.\ref{RNLOPC}, respectively.  

The absolute value of the NLO corrections  do not exceed $6\%$ in the accessible interval of $-t=2.5-6.5$~GeV$^{2}$.   
  For the small values of the momentum transfer the  RCs are negative but they change the sign around $-t=3.5-4$~GeV$^{2}$ and became positive.  
   The  largest effect from  the radiative corrections is observed
for the boundary values  $-t=2.5$~GeV$^{2}$ and $-t=6.46$~GeV$^{2}$.   In this case the NLO contributions for massless case 
 reduce the value of $|\mathcal{R}|$ by  $5\%$  for   $s=10.9$~GeV$^{2}$. 
 This leads to a bit larger sensitivity of the  $|\mathcal{R}|_{NLO}$  to the variable  $s$  at $t=-2.5$~GeV$^{2}$  comparing to
$|\mathcal{R}|_{LO}$.  The  effect from the RCs for the $\bar{\mathcal{R}}$ is quite similar.  Taking into account that the hard-spectator contribution provides of
about $10\%$ of the cross section \cite{Kronfeld:1991kp,Vanderhaeghen:1997my, Brooks:2000nb, Thomson:2006ny} 
we can conclude that computed NLO corrections provide  a  comparable  numerical effect. 
 
The inclusion of the power corrections as described above reduce the absolute value of $\bar{\mathcal{R}}$ in the interval  $0-13\%$ for the different values of $-t$. 
One can also observe  that  the extracted  values $\bar{\mathcal{R}}$  are less sensitive to the value of $s$  than  ${\mathcal{R}}$.  This may  indicate  that the more sensitive  
 to $s$ behavior of ${\mathcal{R}}$ can be associated  with the power corrections. 

\begin{figure}[h]
\begin{center}
\begin{tabular}[c]{ll}
{\includegraphics[width=3.5in]{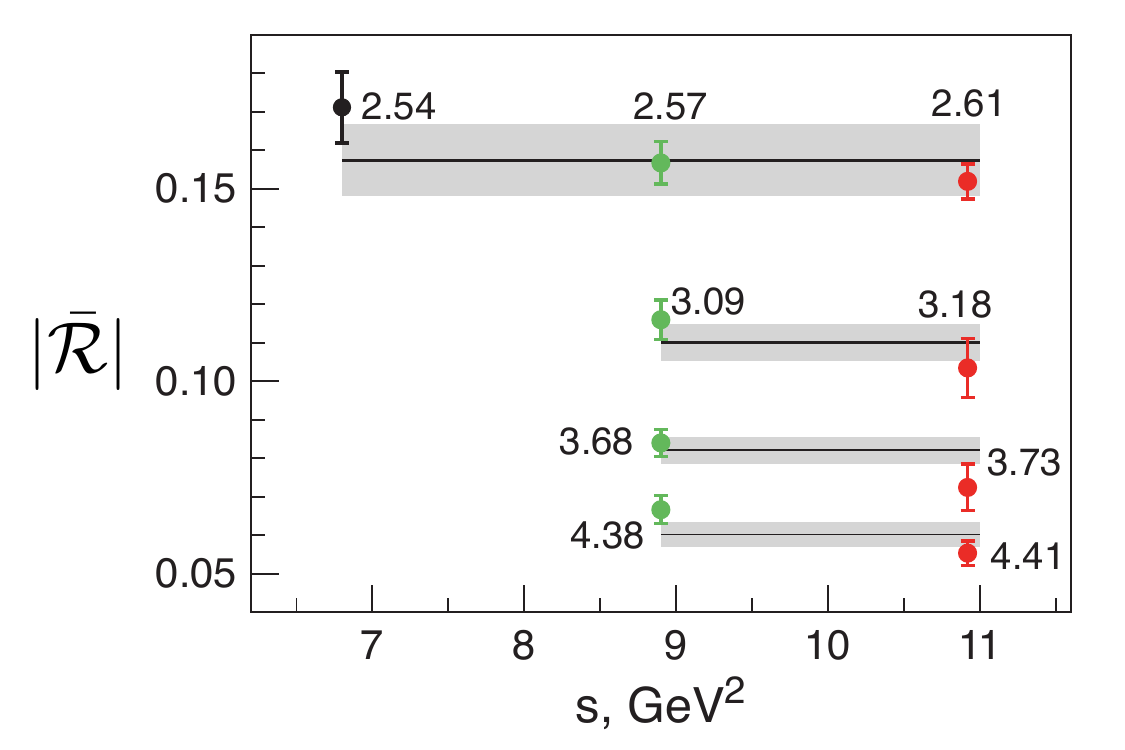}} 
\end{tabular}
\end{center}
\caption{The ratio $\bar{\mathcal{R}}$ as a function of 
 energy $s$ at fixed values of $t$. The fit and bounds are the same as on the right plot in Fig.\ref{RNLOPC}. }
\label{RNLOPC-s}
\end{figure}

 In the  figures describing the ratio ${\mathcal{R}}$  we show the empirical  fit of
the extracted points (solid line) together with the 99\% confidence
bands (gray shaded area). For the empirical fit we used a simple power function
\begin{equation}
|\mathcal{R}(s,t)|=\left(  \frac{\Lambda^{2}}{-t}\right)  ^{\alpha}.
\label{Rfit}%
\end{equation}
where  $\alpha$ and $\Lambda$  are unknown fitting parameters.  The results of the fit for different cases are shown in Table~\ref{fitout}.  One can see  there
that $\chi^{2}$/d.o.f  is much better for $\bar{\mathcal{R}}$  extracted  with the kinematical power corrections.   
This is the consequence  of the less sensitive  behavior of the extracted points for $\bar{\mathcal{R}}$  with respect to energy $s$ as we discussed above.    
\begin{table}[h]
\caption{Results for the parameters $\Lambda$ and $\alpha$  defining  the  behavior  (\ref{Rfit})  for the ratios  $|\mathcal{R}|$ in Fig.\ref{Rnlo} 
and $|\bar{\mathcal{R}}|$ in Fig.\ref{RNLOPC}    }
\label{fitout}
\begin{center}%
\begin{tabular}
[c]{|c|c|c|c|}\hline
\phantom{empty}& {$\Lambda$, GeV} & {$\alpha$} & {$\chi^{2}$/d.o.f } \\ \hline
& & & \\[-3mm] 
$|\mathcal{R}|$, NLO  & $0.95\pm 0.02$ & $1.67\pm 0.05$ &  $2.7$    \\[1mm]\hline 
& & & \\[-3mm] 
  $|\bar{\mathcal{R}}|$, LO  & {$1.0\pm 0.02$} & {$1.88\pm0.05$} & {$1.1$}  \\[1mm]\hline
& & & \\[-3mm] 
 $|\bar{\mathcal{R}}|$, NLO & {$0.98\pm0.02$} & {$1.80\pm 0.05 $} & {$1.25$} \\[1mm]\hline
\end{tabular}
\end{center}
\end{table}
It is also interesting to note that obtained  results for the exponent $\alpha$  are  somewhat  smaller than  the expected asymptotic
power behavior obtained from  the SCET analysis:
$|\mathcal{R}(s,t)|\sim {(-t)}^{-2}$.   But for the discussed  values of the momentum transfer $-t\simeq 2.5-7$GeV$^{2}$ the hard-collinear 
scale $\mu_{hc}\simeq \sqrt{\Lambda Q}$ is still quite small.   Therefore we expect  that these empirical  values of $\alpha$ can be a result  
of the oversimplified choice of the fit formula in Eq.(\ref{Rfit}).   The measurements of the cross section  for the higher values of $-t$ can help to clarify this situation.
In Fig.\ref{RNLOPC-s}  we show the ratio $\bar{\mathcal{R}}$ as a function of 
 energy $s$ at fixed values of $t$ and compare with the obtained fit.    

The other measured  observables  which are very helpful in order  to clarify  the   underlying
partonic dynamics  are given by the  recoil polarizations  $\KLL$ and $\KLS$.   They  can be
constructed for the circular polarized photon ($R,L$) and longitudinal
($\Vert$)  or transverse ($\bot$) polarization of the recoiling proton. 
In the current work we consider only the longitudinal polarization $\KLL$  because it 
does not depend on the helicity flip  amplitudes in the leading power approximation.  
Its  definition reads
\begin{equation}
\KLL =\frac{\sigma_{\Vert}^{R}-\sigma_{\Vert}^{L}}{\sigma_{\Vert}^{R}%
+\sigma_{\Vert}^{L}}.
\label{KLLdef}
\end{equation}
Computing this asymmetry with the help of the  approximation  Eq.(\ref{Ti-ssp})  we obtain
that the unknown factor $|R|$ cancels in the ratio and the asymmetry is defined  only
by  the perturbative coefficients $C_{i}$.

 Neglecting  all power corrections and using the next-to-leading
expressions we obtain
\begin{align}
\KLL  &  =\frac{s^{2}-u^{2}}{s^{2}+u^{2}}-\frac{\alpha_{s}}{\pi}C_{F}%
\frac{1}{\left(  s^{2}+u^{2}\right)  ^{2}}\left\{  ~(t-s)u^{3}\ln^{2}\left[
|u|/|t|\right]  -(t-u)s^{3}\ln^{2}\left[  s/|t|\right]  \right. \nonumber\\
&  \left.  +su^{2}(2t-s)\ln\left[  |u|/|t|\right]  -us^{2}(2t-u)\ln\left[
s/|t|\right]  -\pi^{2}(s-t)u^{3}~\right\}  +\mathcal{O}(\alpha_{s}^{2}),
\label{KLL}%
\end{align}%
The leading-order contribution in  this  expression  reproduces  
the well-known expression for the Klein-Nishina asymmetry  which describes the scattering on the
point-like massless particles.  Obviously, this term does not depend on energy $s$ if we rewrite $u$ in terms of $s$
and scattering angle $\theta$ in the massless approximation.   The weak logarithmic $s$-dependence is introduced by the QCD
radiative correction through the definition of the scale for the running coupling $\alpha_{s}$.

\begin{figure}[h]
\begin{center}%
\begin{tabular}[c]{ll}%
\includegraphics[width=2.8in]{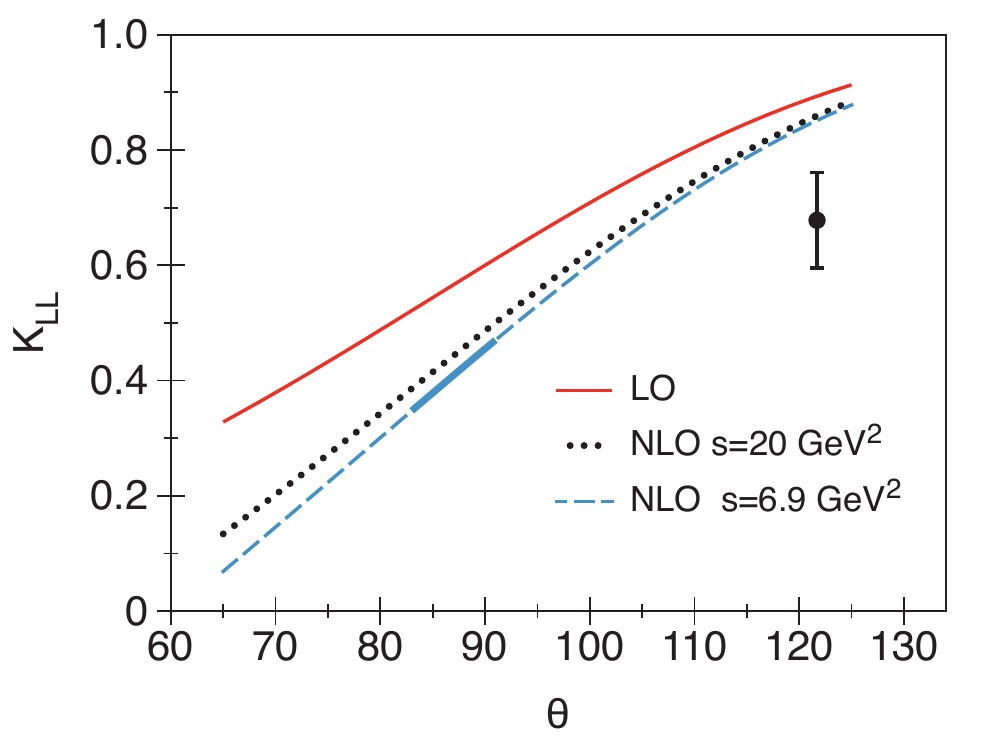} 
& \includegraphics[width=2.8in]{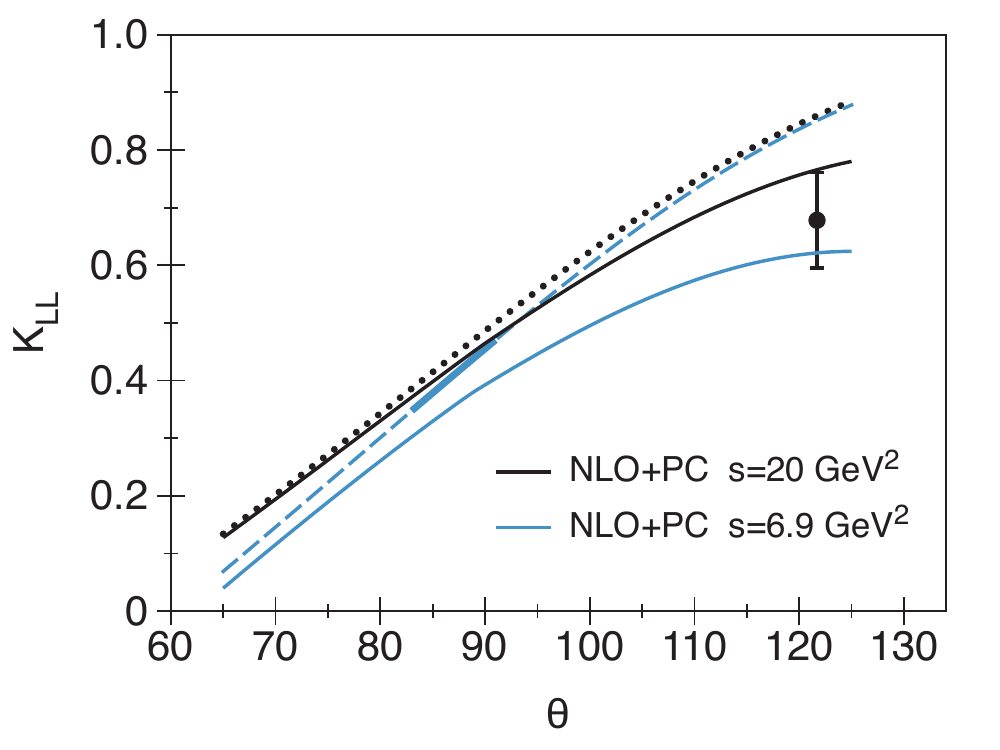}
\end{tabular}
\end{center}
\caption{The longitudinal asymmetry  $\KLL$  as a function of
scattering angle $\theta$. {\bf Left plot:}  comparison of the LO (red solid) and NLO results computed with $s=6.9 , 20$~GeV$^{2}$, (dashed and dotted lines, respectively).  
The kinematical  power corrections (PC) are neglected  $m=0$. 
{\bf Right plot:} comparison of the NLO results  computed with (solid black  and blue lines) and without kinematical power corrections. 
The curves for the massless approximation are the same as on the left plot.  }
\label{asym-fig}%
\end{figure}

In Fig.\ref{asym-fig} (left plot)  we show the numerical results for the  asymmetry $\KLL $ as a functions of the scattering angle $\theta$. 
The solid red line corresponds to the  leading-order approximation in Eq.(\ref{KLL}) (massless Klein-Nishina  asymmetry). 
The dashed (blue) and  dotted (black) lines show the numerical results for the complete NLO expression (\ref{KLL}) for  the energies  $s=6.9$~GeV$^{2}$ and
$s=20$~GeV$^{2}$, respectively. The data point corresponds to  $s=6.9$~GeV$^{2}$ \cite{Hamilton:2004fq}.  We see that the energy dependence of the  NLO expression
remains quite  weak.  The estimates which are obtained  are larger than the experimental data point. However  for this energy  and angle $\theta=121.6^{o}$ the 
value of the $-u=1.14$~GeV$^{2}$  is still quite small.   For clarity we show with the help of the solid thick (blue) line the values of $\KLL$ in the kinematical  interval  
 where  $-t\geq 2.5$~GeV$^{2}$ and  $-u\geq 2.5$~GeV$^{2}$  for $s=6.9$~GeV$^{2}$.  Keeping in mind the  estimates  of the cross section  
 one can expect that the  power corrections  for this kinematical region  can still provide a sizeble numerical effect.   
 
 Therefore in order to estimate  the possible effect from the power suppressed  contributions  we include into consideration  the 
 kinematical power corrections in the same way as  we did for the cross section before.  The  numerical results  are presented  in  Fig.\ref{asym-fig} (right plot).  
 One can observe that the computed power corrections  provide a sizeble effect for large angles ($\theta>90^{o}$) and quite small for the  $\theta\leq 90^{o}$. 
 We see  that their effect for  energy  $s=6.9$~GeV$^{2}$ is  quite  large and negative  bringing the curve in agreement with the data point.  
One can also observe  that  their effect for  large energy $s=20$~GeV$^{2}$ and large angle $\theta=120^{o}$ ($-u=4.5$~GeV$^{2}$) is approximately a factor of three  smaller 
but is still  quite sizeble numerically.

\section{Discussion}
\label{discussion}
We provided  a detailed consideration  of the QCD factorization for the WACS process.  Using the SCET framework we proved   that the  leading-power or
dominant contribution  is described by the soft- and hard-spectator scattering.  For  asymptotically large values of the Mandelstam variables the 
soft-spectator contribution is strongly suppressed by the Sudakov logarithms but not by  powers of a generic large scale $Q$.  In the region of moderate 
values of $Q^{2}$ where the hard-collinear scale $\mu_{hc}\sim \sqrt{Q\Lambda}$ is still quite small  this logarithmic suppression is weak and therefore one 
must include the  soft-spectator contribution on the same footing  as the hard one.  We provided the factorization formulas for the three amplitudes describing    
the scattering when the nucleon helicity is conserved.  The  amplitudes corresponding to the
 helicity flip scattering  are suppressed as $1/Q$ relative to the helicity conserving ones. In the present work, we did not include  these subleading amplitudes  in
the current  analysis.  

In the SCET framework the soft-spectator contribution is described only by one SCET-I  form factor  $\mathcal{F}_{1}$ which must be considered as a non-perturbative function 
in the region of  moderate values of   $Q^{2}$. This form factor depends only on the momentum transfer $t$  due to the underlying hard-collinear scattering. 
 This feature allows one to define the  $s$-behavior of the amplitudes  computing the hard coefficient functions.  
 The tree level hard coefficient functions to the soft-spectator part have been computed in Ref.\cite{Kivel:2012vs}.
 Here we also presented the next-to-leading QCD corrections to these quantities.  This calculation provide a direct check  of the factorization 
 for the soft-spectator part  beyond the tree approximation.    
  
Unfortunately the SCET-I  form factor $\mathcal{F}_{1}$  is  sensitive to  the overlap between the soft and collinear regions.  
This can be seen explicitly when  one  performs the factorization of the hard-collinear modes.  
The same feature also leads  to the end-point singularities  in the hard-spectator contribution, see {\it e.g.}  \cite{Kivel:2012mf}.  
Hence one has to imply a certain rapidity regularization in order to  unambiguously define  the  soft-  and hard-spectator contributions. 
Then the form factor $\mathcal{F}_{1}$  and the hard-spectator contribution also  depend on the  factorization scale associated with the rapidity regularization.    
In order to avoid this difficulty we 
 used the  physical subtraction scheme   and  excluded  the form factor $\mathcal{F}_{1}$  rewriting it in terms of  one physical amplitude. 
 This allows us  to obtain the  well defined relations  Eq.(\ref{T24R})  between  all three dominant amplitudes.  The unknown nonperturbative  dynamics describing the soft-overlap 
 configuration  is included in the ratio $\mathcal R(s,t)$ defined in Eq.(\ref{Rdef}).  This quantity  can be extracted from the analysis of existing  data. 
 
 In the phenomenological analysis these  relations can be further simplified if  the hard-spectator contribution is quite small in the region of  moderate values of $Q^{2}$ and therefore can be neglected.  Such an assumption looks to be   reliable because the hard-spectator coefficient  functions are suppressed as  $\mathcal{O}(\alpha^{2}_{s})$. In addition,  the existing  
  calculations demonstrate that the hard-spectator mechanism predicts cross section which are one order of magnitude smaller  than the experimental data  
  \cite{Kronfeld:1991kp,Vanderhaeghen:1997my, Brooks:2000nb, Thomson:2006ny}.   
  
  The dominance of the soft-spectator configuration  can be checked  through the $s$-dependence  of the  ratio   $\mathcal{R}(s,t)$. If the contribution of the hard-spectator  mechanism is relatively small then the bulk of this dependence is described by the hard-coefficient function in  the soft-spectator term and must cancel according to Eq.(\ref{Rdef}).  We extracted the 
  value of  the ratio $\mathcal{R}(s,t)$ using the existing data  for different  values of  $s$ and $t$.  We  performed  the leading order and the next-to-leading order  analysis 
  and also  investigated the effect of the kinematical power corrections.  The main qualitative conclusion is that  for  the region where $|t|,|u|\geq 2.5$~GeV$^{2}$  a weak 
  $s$-dependence is observed.   In Fig.\ref{Rnlo}  one can see that the data for the same $t$   and different $s$ differ  by $15-20\%$. However, the existing data cover the region of relatively 
  small values of $s$ and $t$.  They can however still be affected from power corrections.  In order to estimate their effect we included the kinematical power corrections which describe the simple powers of $m/Q$ arising  from the exact hadronic kinematics in the expression for the cross section.  In this case the ratio $\mathcal R(s,t)$ is less sensitive to the  energy $s$ as one can see in Fig.\ref{RNLOPC}.  The difference between the extracted values of $\mathcal {R}(s_{i},t)$  are around  $11-13\%$. These results allow us to  conclude that existing   data  are in a good  agreement with the assumption about the dominance of the soft-spectator contribution. 
  
Further  measurements  of the cross section for higher values of the Mandelstam variables can be very helpful  in order to continue to study and to constrain the ratio 
  $\mathcal{R}(s,t)$.  The higher values $s$, $t$ and $u$   will allow one  to reduce the effect  of the power corrections  and  to investigate much  better the role of the different logarithmic QCD corrections. We suppose that the hard-spectator corrections which can provide  the effect  of order $10\%$  must also be included in the phenomenological analysis.  
  
 The factorization framework allows one to use the extracted ratio $\mathcal{R}$  for the analysis of the other processes. 
  In Ref.\cite{Kivel:2012vs} it was shown that  this function also describes the soft-spectator contribution in  the  two-photon exchange (TPE)
  correction  in the elastic lepton proton scattering.  The future  experiments will allow to measure the elastic cross section  
   up to $Q^{2}=17$~GeV$^{2}$. Therefore  in order to reduce the ambiguity in the extraction of the electromagnetic  form factors one need to reduce  the ambiguity from  TPE correction.  
 Therefore we suppose that  the WACS provide us   the best possibility to obtain the required  information  about  this nonperturbative function.

 The different WACS asymmetries can also provide an effective way to study  the detailed underlying hadron dynamics.  We demonstrated that in case of the dominance of the 
 soft-spectator contribution  the longitudinal polarization $\KLL$  can be computed in terms of the hard coefficient functions. 
  Unfortunately the existing data point corresponds to a very low value of $-u=1.14$~GeV$^{2}$.  The inclusion of the kinematical power corrections allows us
   to describe the asymmetry $\KLL$, although    the present treatment of power corrections is mainly meant to indicate the sensitivity of the present data to such effects. 
    We think that  new data for $\KLL$ at higher values of energy and momentum transfer 
 will be  very useful in order to clarify unambiguously the role of the soft-overlap mechanism.      From the theoretical side it is also important  to clarify the  contribution
 of the helicity flip amplitudes which are required in order to describe the transverse polarization $\KLS$  and  in order to reduce the corresponding ambiguity  in  other observables.

\section*{ Acknowledgments}
 This work was supported by the Helmholtz Institute Mainz.

\section*{Appendix A }
\label{AppA}

\begin{figure}[ptb]
\centering
\includegraphics[width=2.8028in]{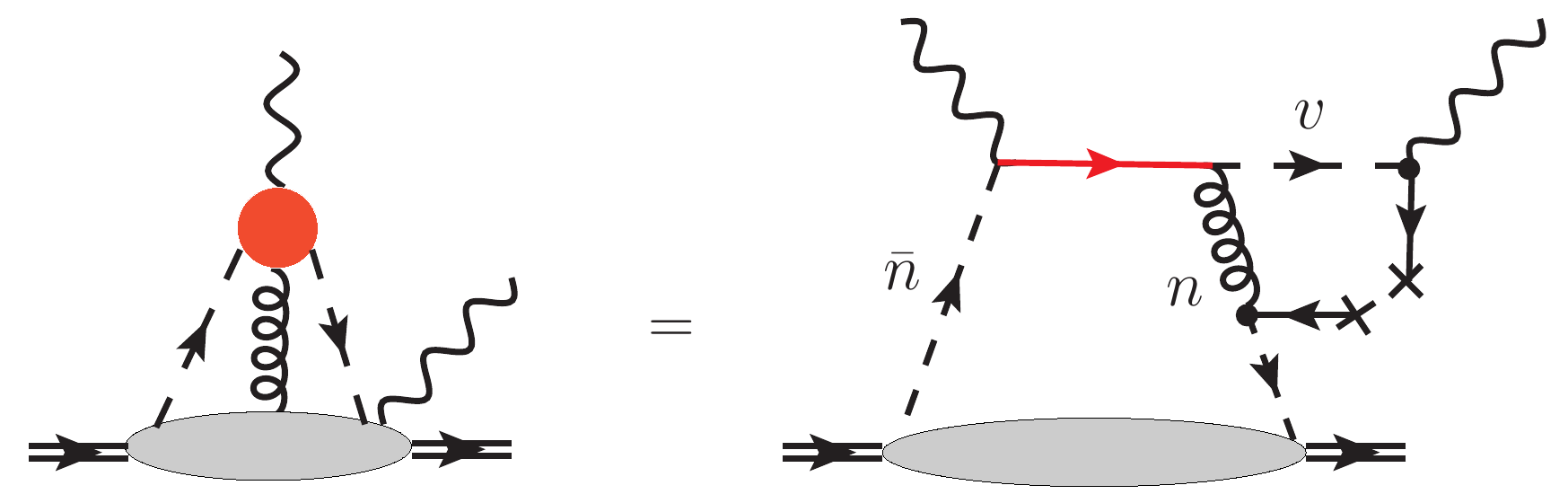}
\caption{The diagrams describing the configuration with the three-particle
soft-overlap described by the SCET operator in Eq.(\ref{MO3jet}). The hard blob is shown by red color. The indices near the
hard-collinear lines show the collinear sector. The soft quarks are shown by
solid lines with the crosses. }
\label{o3jet}
\end{figure}

Here we discuss the soft-overlap contributions involving the photon states. 
The set of the corresponding operators must consist at least from the three
different collinear fields associated with the light-cone vectors $n,\bar{n}$
(nucleon sectors) and $v,\bar{v}$ (photon sectors). Therefore the simplest set
of the operators  reads
\begin{equation}
\tilde{O}^{(3)}=\left\{  \bar{\chi}_{v}\mathcal{A}_{\bot}^{\bar{n}}\chi
_{n},~...\right\}  , \label{O3tld}%
\end{equation}
where dots describe the analogous operators but with the different
hard-collinear fields. The factorization of the amplitude in this case can be described by
the diagrams in Fig.\ref{o3jet}. 
This configuration describes the soft-overlap between the nucleons and outgoing
photon state. One soft quark is attached to the photon vertex. For simplicity,  the
soft-overlap part  associated  with the nucleon sector  is shown by the gray blob.    In this case  the hard-collinear dynamics is
described by the following SCET matrix element
\begin{equation}
\left\langle p^{\prime},q^{\prime}\right\vert \tilde{O}_{1}^{(3)}\left\vert
p\right\rangle _{\text{\tiny SCET}}. \label{MO3jet}%
\end{equation}
We need to  compare  the  contribution of this matrix element  to the contribution of the matrix element in Eq.(\ref{defF1}). 
In order to perform this comparison we proceed in the following way
\bea
&& 
\left\langle p^{\prime},q^{\prime}\right\vert \tilde{O}_{1}^{(3)}\left\vert
p\right\rangle _{\text{\tiny SCET}}=\left\langle p^{\prime},q^{\prime}\right\vert T\{\tilde{O}_{1}^{(3)},
\mathcal{L}_{\text{int}}^{(2,v)}[\bar{q}B^{c}_{\bot}\xi] ,
\mathcal{L}_{\text{int}}^{(1,n)}[\bar{\xi}A_{\bot}q ]  \}
\left\vert  p\right\rangle _{\text{\tiny SCET}}
\\  &&
= ie e_{q}\varepsilon^{*}_{\mu}
\left\langle p^{\prime}\right\vert \int d^{4}x\  e^{i(q'x)}  T\{\tilde{O}_{1}^{(3)},\ \bar q(x_{-})\gamma^{\mu}\xi_{v}(x),\mathcal{L}_{\text{int}}^{(1,n)}[\bar{\xi}A_{\bot}q ]  \} 
\left\vert  p\right\rangle _{\text{\tiny SCET}}.
\label{O3tL11}
\eea
Here the  interaction vertex $\mathcal{L}^{(2,v)}_{\text{int}}[\dots]$ describes the interaction of the
collinear photon (field $B^{c}_{\bot}$) with hard-collinear and soft quarks. The interaction 
$\mathcal{L}_{\text{int}}^{(1,n)}[\dots]$  is given in Eq.(\ref{xiAq}).  Computing the  contractions in Eq.(\ref{O3tL11}) one obtains the diagram shown on the {\it rhs} in Fig.\ref{o3jet}.  
In order to perform the complete  matching to SCET-II one must add additional vertices in Eq.(\ref{O3tL11})  which are required for description of 
the hard-collinear interactions associated with the $n$ and $\bar{n}$
collinear sectors.  These insertions can be constructed in the same way as in $T$-product in
Eq.(\ref{TO1q}) in the each collinear sector.   After that one obtains  the following result,  (using the same
notations as in Eq.(\ref{Ok-gen}):
\begin{align}
 \int d^{4}x\  e^{i(q'x)}  T\{\tilde{O}_{1}^{(3)},\ \bar q(x_{-})\gamma^{\mu}\xi_{v}(x),\dots \}
\simeq 
J_{v}^{\gamma}\ast O_{n}^{(6)}\ast J_{n}\ast \tilde O_{S}\ast J_{\bar{n}%
}\ast O_{\bar{n}}^{(6)}, \label{TO3tld}%
\end{align}
where the jet -function $J_{v}^{\gamma}$ describes the hard-collinear
fluctuations associated with the outgoing photon.  

 The relative order  of the matrix elements (\ref{defF1})  and (\ref{MO3jet}) can be obtained  from the estimate of the operator in the matrix element (\ref{O3tL11}). For that 
 operator one obtains
 \be{estO13}
 \int d^{4}x\  e^{i(q'x)}  T\{\tilde{O}_{1}^{(3)},\ \bar q(x_{-})\gamma^{\mu}\xi_{v}(x),\mathcal{L}_{\text{int}}^{(1,n)}[\bar{\xi}A_{\bot}q ]  \}\sim \mathcal{O}(\lambda^{4}).
 \ee
Hence we can conclude that the contribution  (\ref{MO3jet}) is  suppressed as $\mathcal{O}(\lambda^{2})$ compared to  Eq.(\ref{defF1}) and therefore can
be neglected.

\begin{figure}[ptb]
\centering
\includegraphics[width=2.5 in]{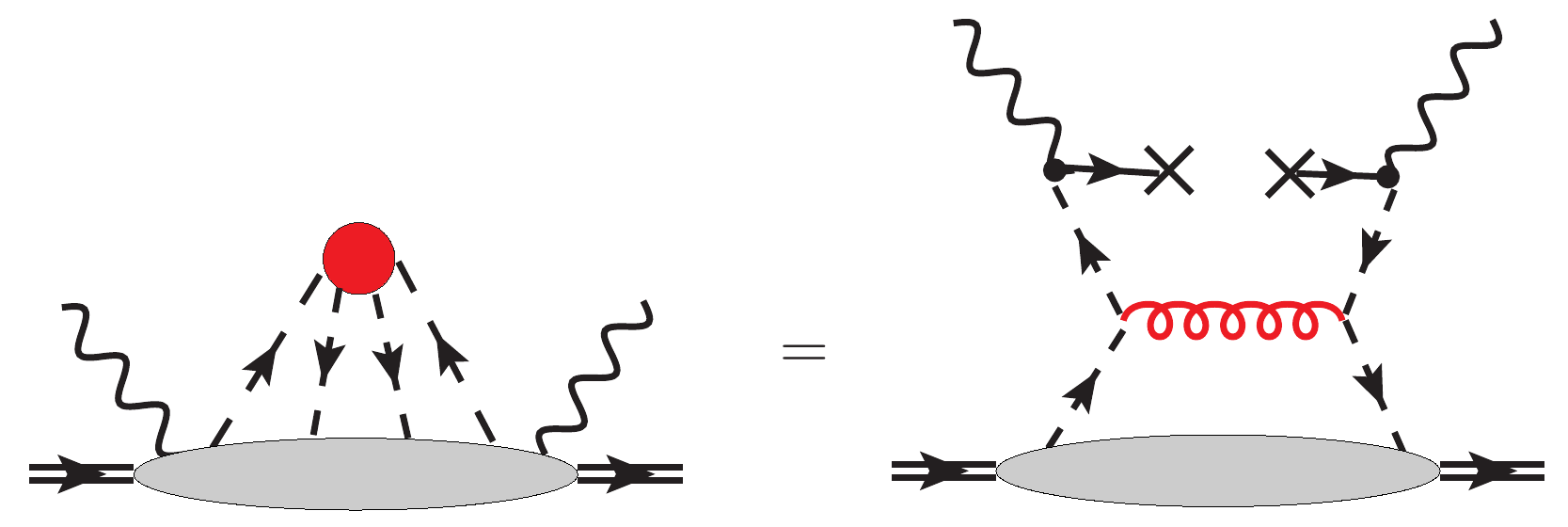}
\caption{An example of the diagram desribing the soft-overlap contribution
with the operator $\tilde{O}^{(4)}$ Eq.(\ref{tldO4}). For simplicity the gray
blob on {\it rhs } denotes the hard-collinear and soft particles which describe the
collinear sectors associated with the nucleon states. The red gluon line denotes the
hard gluon exchange. The interactions of the photons with quarks are described
by the SCET vertices $\mathcal{L}_{\text{int}}^{(2,v)}[\bar{\xi}B^{c}_{\bot}q]$ and $\mathcal{L}_{\text{int}}^{(2,\bar v)}[\dots]$ with the collinear photon fields. }
\label{o4tld}%
\end{figure}

The one more possibility  is provided by  the soft-overlap of the all external states. 
In this case the factorization of the hard modes provides  the four-quark operator  (for simplicity we do not show the indices of the fields):
\begin{equation}
\tilde{O}^{(4)}=\bar{\chi}_{n}\chi_{\bar{n}}~\bar{\chi}_{\bar{v}}\chi_{v}~.
\label{tldO4}
\end{equation}
Such operator arises from the diagram with the hard gluon exchange as shown in
Fig.\ref{o4tld}.
In order to compare this  contribution with the  soft-overlap described  by  FF $\mathcal{F}_{1}$  we have to compare the matrix elements  (\ref{defF1}) with
\bea
&&\left\langle p^{\prime},q^{\prime}\right\vert \tilde{O}^{(4)}\left\vert
p,q \right\rangle _{\text{\tiny SCET}}= 
-e^{2}e^{2}_{q} \varepsilon_{\nu}  \varepsilon^{*}_{\mu} 
\nonumber \\ &&
 \phantom{-e^{2}e^{2}_{q} \varepsilon_{\nu}  \varepsilon^{*}_{\mu}}
 \times \left\langle p^{\prime}\right\vert 
 \int d^{4}x \int d^{4}y\ e^{i(q'x)-i(qy)}
 T\{ 
 \tilde{O}^{(4)},\ \bar \xi_{v}(x)\gamma^{\mu}q, \bar q \gamma^{\nu}\xi_{\bar v}(y)
 \}
 \left\vert p \right\rangle _{\text{\tiny SCET}}, 
 \label{MO4jet}
\eea
where we again do not show the interactions  associated with the nucleon states.  The calculation of the $T$-product in Eq.(\ref{MO4jet}) can 
be illustrated by the diagram on {\it rhs} in Fig.\ref{o4tld}.   The complete  matching of the  operator (\ref{o4tld}) onto  SCET-II can be represented  as
\begin{eqnarray}
&& \int d^{4}x \int d^{4}y\ e^{i(q'x)-i(qy)} T\left\{  \bar{\chi}_{n}\chi_{\bar{n}}
~\bar{\chi}_{\bar{v}}\chi_{v},
\mathcal{L}_{\text{int}}^{(2,v)}[\bar{\xi}B^{c}_{\bot}q],
\mathcal{L}_{\text{int}}^{(2,\bar{v})}[\bar{q}B^{c}_{\bot}\xi],
...\right\}  
\nonumber \\
&&\phantom{empty space }\simeq \left(  J_{v}^{\gamma}\ast J_{\bar{v}}^{\gamma}\right)  \ast
O_{n}^{(6)}\ast J_{n}\ast \tilde O_{S}\ast J_{\bar{n}}\ast O_{\bar{n}}^{(6)},
\end{eqnarray}
where dots denote the insertions from the hard-collinear sectors
associated with the nucleon states. Taking into account that  
\be{estO4t}
\int d^{4}x \int d^{4}y\ e^{i(q'x)-i(qy)}
 T\{ 
 \tilde{O}^{(4)},\ \bar \xi_{v}(x)\gamma^{\mu}q(x), \bar q(y) \gamma^{\nu}\xi_{\bar v}(y) \} \sim 
 \mathcal{O}
(\lambda^{4}),
\ee
we can conclude that the  matrix element in Eq.(\ref{MO4jet}) is suppressed as $\mathcal{O}
(\lambda^{2})$ comparing to the matrix element   (\ref{defF1}).

\begin{figure}[ptb]
\centering
\includegraphics[height=0.787in, width=1.3076in]
{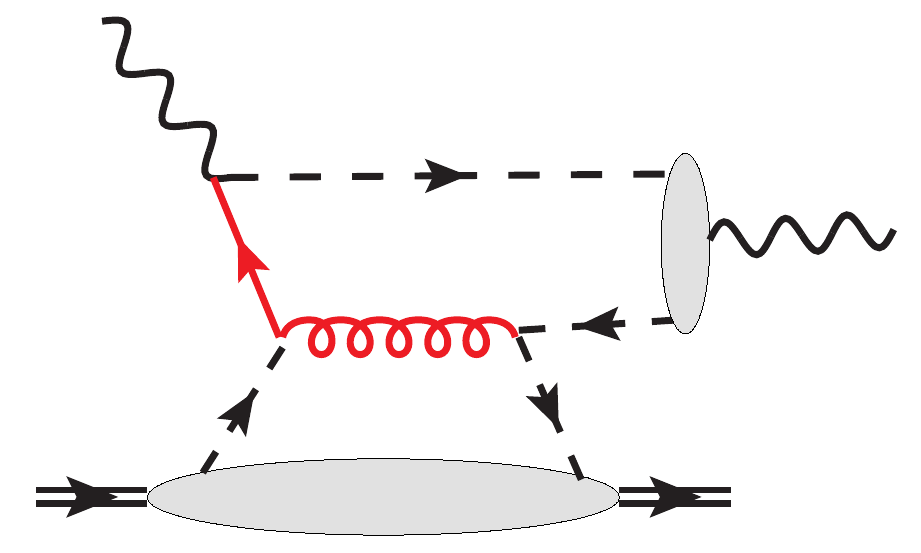}
\caption{The diagram describing the  factorization to the four-quark
operator $O_{\gamma}$ defined in Eq.(\ref{Og}). }
\label{o6g}%
\end{figure}

One more possibility is obtained  when one of the photons  is  considered as a  nonperturbative particle, like a hadron.  
 The corresponding situation  is described by the  operator constructed from the hard-collinear and collinear fields, for instance
\begin{equation}
O^{(6)}_{\gamma}=\bar{\chi}_{n}\chi_{\bar{n}}~\bar{\chi}_{\nu}^{c}\chi_{\nu}^{c},
\label{Og}
\end{equation}
where we do not show the indices and the Dirac structures for simplicity. 
This operator  arises after factorization of the hard modes in the diagram  shown in Fig.\ref{o6g}.
The collinear fields $\bar{\chi}_{\nu}^{c}\chi_{\nu}^{c}$ describe the
nonperturbative overlap of the quarks with the external photon state.
The corresponding matrix element is known as a photon distribution amplitude
\cite{Balitsky:1997wi,Balitsky:1989ry}. However one can easily see that  the
photon matrix element introduces extra suppression  factor $\lambda^{2}$
\begin{equation}
~\left\langle p^{\prime},q^{\prime}\right\vert \bar{\chi}_{n}\chi_{\bar{n}%
}~\bar{\chi}_{\nu}^{c}\chi_{\nu}^{c}\left\vert p\right\rangle _{\text{\tiny SCET}%
}\simeq\left\langle p^{\prime}\right\vert \bar{\chi}_{n}\chi_{\bar{n}%
}\left\vert p\right\rangle _{\text{\tiny SCET}}\left\langle q^{\prime}\right\vert
\bar{\chi}_{\nu}^{c}\chi_{\nu}^{c}\left\vert 0\right\rangle _{\text{\tiny SCET}%
}\sim\lambda^{2}\left\langle p^{\prime}\right\vert \bar{\chi}_{n}\chi_{\bar
{n}}\left\vert p\right\rangle _{\text{\tiny SCET}}\text{.}%
\end{equation}
Hence  this  configuration is also subleading.

\section*{Appendix B}
\label{appendix}
In this section we provide  a detailed discussion of the calculation of  the {\it rhs} of Eq.(\ref{Cnlodif}). 
 The off-shell prescription which is used for the regularizations of IR-divergencies
introduces the soft scale explicitly and this feature  complicates the computation of the
integrals for the  matrix  elements.   One can simplify the straightforward  calculation of  required  integrals 
using  that only  the difference on the \textit{rhs} in Eq.(\ref{Cnlodif}) is needed.
It is clear that  if the factorization theorem is valid then the soft scales must cancel in the difference. 
One can observe this cancellation already at the level of the regularized
integrals using  the technique known as expansion by regions, see {\it e.g.} Refs.\cite{Beneke:1997zp, Smirnov:2002pj}. 

In this approach  the required  integrals  can be represented
as a sum of the contributions associated with the hard,  collinear and soft
regions. The  hard integrals  does not depend on  the soft scales because we need only the leading power contribution.
 Hence only  the collinear and soft contributions can depend on the soft scales $p^{2}$ and $p'^{2}$.  
 The contribution from the soft region arises  only in the 
 box diagram  $D^{\mu\nu}_{c}$  in Fig.\ref{nlo-digrams}  and  in the  crossed  box  $\bar D^{\mu\nu}_{c}$. 
 The contributions from the collinear regions are provided  by  the box diagram $D^{\mu\nu}_{c}$,   vertex diagrams $D^{\mu\nu}_{a,b}$ and the crossed analogs  $\bar D^{\mu\nu}_{a,b,c}$.   
 The self-energy contributions  $D^{\mu\nu}_{d},  \bar D^{\mu\nu}_{d}$  have UV-divergencies and  can be easily computed within the standard technique.  
 Computing the quark wave function renormalizations  we always apply  the on-shell subtractions.  
 We  verified that on the {\it rhs} of Eq.(\ref{Cnlodif})  the soft contribution from the box diagram cancel  with appropriate contribution from the unrenormalized diagram $(c)$ in 
 Fig.\ref{nlo-scet}.  Similarly,  the collinear contributions associated with the quark momenta $p$ and $p'$  cancel 
  with the unrenormalized contributions  provided  by the diagrams $(a,b)$ in Fig.\ref{nlo-scet}.

  In order to be specific  let us  present  the contribution of  the each    QCD diagrams in Fig.\ref{nlo-digrams} as the  following  expression
 \bea
\frac{\alpha_{s}}{\pi}D_{i}^{\mu\nu}=\bar{u}_{n}\gamma_{\bot}^{\sigma}u_{\bar{n}}\ e_{q}^{2}\frac{\alpha_{s}}{\pi}
\sum_{k} N_{k}^{\mu\nu\sigma}(p,p',q)  J_{k}.
\label{defJ}%
\eea
We  imply  that  the expressions  for each  diagram $D^{\mu\nu}_{i}$ can be reduced  to a set of the scalar  integrals $J_{k}$ with the 
coefficients $N_{k}^{\mu\nu\sigma}(p,p',q)$  depending only from the external momenta.  This reduction is carried out in  $D=4$ for the UV-finite expressions. 
The UV-divergent integrals  in the vertex  diagrams  and the self-energy correction  $D^{\mu\nu}_{d}$  are IR-finite and the reduction to the expressions in Eq.(\ref{defJ}) must be carried out  in dimensional regularization (DR).  The corresponding UV-divergent integrals $J_{k}$ can also be easily computed. 
Computing the contributions of the diagrams  $D^{\mu\nu}_{e,f}$ in Fig.\ref{nlo-digrams} we also used DR and subtract the finite piece as required by the  on-shell renormalization prescription. 
   
The most complicated  integrals  are given by the UV-finite and IR-divergent $J_{k}$ which are regularized by the off-shell momenta.   These integrals arise from  the box and 
vertex  diagrams $D^{\mu\nu}_{a,b,c}$.  The dominant regions of integration are described  by  the hard $(h)$,  hard-collinear  or collinear $(c)$ and soft $(s)$ momenta. 
Each scalar integral can be  presented as a sum  of the regularized integrals 
\be{Jksum}
J_{k}=J^{(h)}_{k}+ J^{(c)}_{k}+J^{(s)}_{k}
\ee
where  each integral on the {\it rhs}  is divergent and must be regularized with the help of DR.  Here we assume that collinear regions include  the directions associated with 
all external momenta $p, p', q, q'$.  However we expect that collinear  contributions  associated with the photon momenta  
must cancel  as required by the factorization. 

Using (\ref{defJ})  we rewrite  Eq.(\ref{Cnlodif})  in the  following form
\begin{equation}
C^{\mu\nu\sigma}_{\text{{\tiny NLO}}} =
 \sum \left( J_{\text{reg}}^{\mu\nu\sigma}+
J_{(h)}^{\mu\nu\sigma}+ J_{(n)}^{\mu\nu\sigma}+ J_{(\bar{n})}^{\mu\nu\sigma}
+ J_{(v)}^{\mu\nu\sigma}+ J_{(\bar{v})}^{\mu\nu\sigma}+
J_{(s)}^{\mu\nu\sigma}
\right)
-\ C_{\text{{\tiny LO}}}^{\mu\nu\sigma} \hat{\mathcal{F} }_{NLO}^{q}.
\label{CNLO=sumJ}
\end{equation}
Here the sum denotes the summation over the all diagrams $D_{i}$ and also over the  
integrals  $N_{k}^{\mu\nu\sigma}J_{k}\equiv J^{\mu\nu\sigma} $ which have been expanded according to (\ref{Jksum}). 
The indices $h,n,\bar{n},v,\bar{v},s$ denotes the integrals representing
the hard, collinear to $p^{\prime}$,$p,q^{\prime},q$ and soft regions, respectively.  The term  $J^{\mu\nu\sigma}_{\text{reg}}$  denotes the 
contributions which is obtained from the sum of the  UV-divergent but  IR-finite integrals.  Because the UV-poles cancel this contribution is finite ( or regular). 

Let us consider the  SCET matrix element   describing  the quark form factor   $\hat{\mathcal{F} }^{q}$.  We rewrite it  as multiplicatively renormalized  bare form factor
\be{ZF}
~\hat{\mathcal{F}}^{q}  & =
Z^{-1}_{O} (Z^{\text{\tiny OS}}_{p})^{1/2}(Z^{\text{\tiny OS}}_{p'})^{1/2} \left( \hat{\mathcal{F}}^{q}\right) _{B},
\ee
where the subscript $B$ denotes the unrenormalized or bare form factor, the factors $Z^{\text{\tiny OS}}_{q}$ denote the renormalization constant of the quark field in the on-shell scheme, 
the factor $Z^{-1}_{O}$ describe the $\overline{MS}$ counterterms for the operator vertex.  Remind that the unrenormalized form factor is given by the diagrams in Fig.\ref{nlo-scet}.  Let us  write
\be{ZoNLO}
Z^{-1}_{O} =1-\frac{\alpha_{s}}{\pi}Z^{NLO}_{O} +\dots,  
\ee
where dots denote the higher order contributions. The bare form factor is given by the sum of the diagrams in Fig.\ref{nlo-scet}. We write the contribution of  each diagram as
\be{defDiscet}
D_{i}=\bar{u}_{n}\gamma_{\bot}^{\sigma}u_{\bar{n}}\ e_{q}^{2}\frac{\alpha_{s}}{\pi}\mathcal{D}_{i}
\ee
Now  Eq.(\ref{ZF}) can be presented in the following form
\be{Fq=ZD}
\hat{\mathcal{F}}^{q}_{NLO}  & = \mathcal{D}_{a}+\mathcal{D}_{b}+\mathcal{D}_{c} - Z^{NLO}_{O} \hat{\mathcal{F}}^{q}_{LO}.
\ee 
The  contributions of the self-energy diagrams $\mathcal{D}_{d,e}$  cancel  with  the NLO  terms from $Z^{\text{\tiny OS}}_{q}$.  Substituting  (\ref{Fq=ZD}) into Eq.(\ref{CNLO=sumJ}) 
(remind that $\hat{\mathcal{F}}^{q}_{LO}=1$) yields
\bea
C^{\mu\nu\sigma}_{\text{{\tiny NLO}}} & =&
  \sum  J_{\text{reg}}^{\mu\nu\sigma}
 + 
 \sum J_{(h)}^{\mu\nu\sigma} -C_{\text{{\tiny LO}}}^{\mu\nu\sigma}  Z^{NLO}_{O}
 \nonumber \\ &&
 +  \sum \left(
J_{(v)}^{\mu\nu\sigma}+ J_{(\bar{v})}^{\mu\nu\sigma}
\right)
\nonumber \\  &&
+  \sum \left(
 J_{(n)}^{\mu\nu\sigma}+ J_{(\bar{n})}^{\mu\nu\sigma}
 \right)- C_{\text{{\tiny LO}}}^{\mu\nu\sigma} (\mathcal{D}_{a}+\mathcal{D}_{b}) 
\nonumber \\  &&
 +
\sum  J_{(s)}^{\mu\nu\sigma}-C_{\text{{\tiny LO}}}^{\mu\nu\sigma} \mathcal{D}_{c} 
\label{Chsc}
\eea
In the last two lines of this equation we grouped the contributions which depend on the soft scales $p^{2}$ or $p'^{2}$.  These  lines include  the soft and collinear  contributions.  Performing the comparison of the corresponding integrands one can observe a lot of cancellations without  the explicit  computation  of the corresponding integrals.  In general case these differences can provide  simple scale independent terms. In the present the  various terms  in Eq.(\ref{Chsc}) cancel exactly:
\bea
 \sum \left( J_{(n)}^{\mu\nu\sigma}+ J_{(\bar{n})}^{\mu\nu\sigma} \right)- C_{\text{{\tiny LO}}}^{\mu\nu\sigma} (\mathcal{D}_{a}+\mathcal{D}_{b})=0,\\
\sum  J_{(s)}^{\mu\nu\sigma}-C_{\text{{\tiny LO}}}^{\mu\nu\sigma} \mathcal{D}_{c} =0.
\eea
The contributions associated with the collinear to $v$ and $\bar v $ regions must also cancel  otherwise they violate the factorization theorem.  
Therefore using this  we obtain 
\be{Cnlo-fin}
C^{\mu\nu\sigma}_{\text{{\tiny NLO}}}  =
  \sum  J_{\text{reg}}^{\mu\nu\sigma} + \sum J_{(h)}^{\mu\nu\sigma} - Z^{NLO}_{O} C_{\text{{\tiny LO}}}^{\mu\nu\sigma}.
 \ee
 The expression for the $Z^{NLO}_{O}$ reads
 \be{ZNLO}
  \frac{\alpha_{s}}{\pi}Z^{NLO}_{O} =- \frac{\alpha_{s}}{4\pi}%
C_{F}\left(  \frac{2}{\varepsilon^{2}}-\frac
{2}{\varepsilon}\ln[-t/\mu^{2}]+\frac{4}{\varepsilon}\right).
 \ee
These poles cancel the UV-divergencies  in  the diagrams  $\mathcal{D}_{a,b,c}$ in Fig.\ref{nlo-scet}.   In   Eq.(\ref{Cnlo-fin})  the poles $Z^{NLO}_{O}$ can be compensated
only by the IR-poles from the hard integrals $J_{(h)}$.  This balance provides a powerful check of the factorization theorem.  Let us also remind that the hard integrals  $J_{(h)}^{\mu\nu\sigma}$ originate from the  QCD diagrams $D^{\mu\nu}_{a,b,c}$ in Fig.\ref{nlo-digrams}.  Performing the evaluation of the {\it rhs}  in Eq.(\ref{Cnlo-fin}) we also checked the validity of the electromagnetic  gauge invariance: $q_{\nu}C^{\mu\nu\sigma}_{\text{{\tiny NLO}}}=q'_{\mu}C^{\mu\nu\sigma}_{\text{{\tiny NLO}}}=0.$   The  results for the scalar coefficients  $C_{2,4,6}^{\text{{\tiny NLO}}}$   given in Eqs.(\ref{C2NLO} -\ref{C6NLO})  were obtained from the   Eq.(\ref{Cnlo-fin})  performing the projections on the scalar amplitudes $T_{2,4,6}$  with the help of  the tensor structures  (\ref{T1-6})  
\bea
-T_{12}^{\mu\nu}\sum \frac{\alpha_{s}}{\pi}( J_{\text{reg}}^{\mu\nu\sigma} +  J_{(h)}^{\mu\nu\sigma} - Z^{NLO}_{O} C_{\text{{\tiny LO}}}^{\mu\nu\sigma})
=q^{\sigma}_{\bot}\frac{\alpha_{s}}{4\pi}C_{F}  C_{2}^{\text{{\tiny NLO}}}(s,t;\mu^{2})  ,\label{T12NLO}
\\
-T_{34}^{\mu\nu}\sum \frac{\alpha_{s}}{\pi}( J_{\text{reg}}^{\mu\nu\sigma} +  J_{(h)}^{\mu\nu\sigma}- Z^{NLO}_{O} C_{\text{{\tiny LO}}}^{\mu\nu\sigma}) 
=q^{\sigma}_{\bot}\frac{\alpha_{s}}{4\pi}C_{F} C_{4}^{\text{{\tiny NLO}}}(s,t;\mu^{2}) ,\label{T34NLO}
\\
\frac{1}{2}T_{6}^{\mu\nu}\sum \frac{\alpha_{s}}{\pi}( J_{\text{reg}}^{\mu\nu\sigma} +  J_{(h)}^{\mu\nu\sigma}- Z^{NLO}_{O} C_{\text{{\tiny LO}}}^{\mu\nu\sigma} ) =
q^{\sigma}_{\bot}\frac{\alpha_{s}}{4\pi}C_{F}
 C_{6}^{\text{{\tiny NLO}}}(s,t;\mu^{2}) .  \label{T6NLO}
\eea
Computing $C_{i}^{\text{{\tiny NLO}}}$  we obtained that  the $1/\varepsilon$ poles on the {\it lhs} of Eqs.(\ref{T12NLO}-\ref{T6NLO})  cancel  as it  is required by factorization.  
The  $\mu$-dependence  of the coefficient functions $C_{2,4,6}^{\text{{\tiny NLO}}}$  can be also checked with the 
help of the RG-equation as discussed in the Sec.\ref{calculation}.


\end{document}